\documentclass[10pt,english]{article} 

\usepackage{amsmath, amsthm, amssymb}
\usepackage{a4, babel, graphicx, listings, fancyhdr, verbatim}
\usepackage{mathrsfs} 
\usepackage{anysize} 
\usepackage{natbib} 
\usepackage[format = hang]{caption}                              

\usepackage{a4}
\usepackage[latin1]{inputenc}
\usepackage{amsmath, amsthm}
\usepackage{graphicx}
\usepackage{amssymb}
\usepackage{verbatim}
\usepackage{listings}
\usepackage{alltt}
\usepackage{babel}
\usepackage{mathrsfs}
\usepackage{ae}
\usepackage{natbib}
\usepackage{booktabs} 

\newtheoremstyle{theorem}
{10pt} 
{10pt} 
{\sl} 
{\parindent} 
{\bf} 
{. } 
{ } 
{} 

\theoremstyle{theorem}

\def\new{{\rm new}}

\def\beq{\begin{eqnarray}}
\def\eeq{\end{eqnarray}}

\def\beqn{\begin{eqnarray*}}  
\def\eeqn{\end{eqnarray*}}

\def\Var{{\rm Var}}

\def\cov{{\rm cov}}

\def\N{{\rm N}}

\def\Pr{P}

\def\hatt{\widehat}

\def\sumin{\sum_{i=1}^n}
\def\sumjk{\sum_{j=1}^k}

\def\eps{\varepsilon}
\def\half{\hbox{$1\over2$}}

\def\rootn{\sqrt{n}}
\def\data{{\rm data}}
\def\obs{{\rm obs}}

\def\tr{{\rm t}}

\def\prof{{\rm prof}}

\def\sd{{\rm sd}}

\def\aic{{\rm aic}}

\def\pois{{\rm Pois}}

\def\cc{{\rm cc}}

\numberwithin{equation}{section} 
\numberwithin{figure}{section}
\numberwithin{table}{section}
\usepackage{hyperref}

\title{The stochastic view used in climate sciences:
   (some) perspectives from (some of)
   mathematical statistics}

\def\idag{{November 2025}}
\date{\idag}

\begin{document}


\maketitle

\centerline{\large\bf Nils Lid Hjort}


\medskip 
\centerline{\bf Department of Mathematics, University of Oslo}

\begin{abstract}
\noindent
Climate statistics is of course a very broad field,
along with the many connections and impacts for yet other areas,
with a history as long as mankind has been
recording temperatures, describing drastic weather events, etc.
The important work of Klaus Hasselmann,
with crucial contributions to the field,
along with various other connected strands of work, 
is being reviewed and discussed in other chapters. 
The aim of the present chapter is to point to a few
statistical methodology themes of relevance for
and joint interest with climate statistics.
These themes, presented from a statistical methods perspective, include
(i) more careful modelling and model selection
strategies for meteorological type time series;
(ii) methods for prediction, not only for future
values of a time series, but for assessing when
a trend might be crossing a barrier, along with
relevant measures of uncertainty for these; 
(iii) climatic influence on marine biology;
(iv) monitoring processes to assess whether and then
to what extent models and their parameters have stayed
reasonably constant over time;
(v) combination of outputs from different information sources; and
(vi) analysing probabilities and their uncertainties
related to extreme events. 
\end{abstract}


\section{Introduction}
\label{section:intro}

Broadly speaking, there has been and is a fairly healthy cooperation
between (i) climate statistics, theoretical and applied, and
(ii) what might be called statistical modelling and methodology.
Connection points for this `stochastic view'
include basic regression modelling,
time series methodology, tools of multivariate statistics,
elements of statistical causality, prediction, 
attempts at fusing physics with imperfect measurements,
distinguishing signal from noise, etc.
Yet there might be other application domains where
the interplay between direct science and statistics theory
development has been or is stronger and more symbiotic.
Instances of such heftier cooperation, so to speak,
might be the many themes associated with medicine,
health care, bioinformatics, the voluminous oil and gas industry,
the intricacies and practicalities of international power grids,
mathematical finance, and the various modern technological inventions
involving machine learning, the omnipresent
Large Language Models, and AI in general. 

It is with such perspectives in mind, coming myself from
the camp of statistics methodology, with experiences
from various points of contact with applied sciences,
and after friendly encounters with {\it some} climate
statistics scholars related to {\it some} of their themes, 
that I shall attempt to point to the potential relevance
of {\it some} methodological developments. There is certainly
room for more active collaboration, which I would very much
welcome. 

To the extent that one may identify `two cultures', in a Snowian fashion, 
the climate statistics scholars and the perhaps more university
professor dominated methodologically oriented statisticians,
it is of course hard to define or estimate any meaningful
Mahalanobis distance between them, and the associated confidence
intervals would be wide. The two camps will naturally 
have considerable overlap, both regarding scholars with
a foot in both worlds and regarding themes worked with. 
It might nevertheless be useful to put up some key words
and phrases for the two.

The climate statisticians are applied researchers working
at the interface of data, models, and decision-making,
perhaps also touching politics. Their primary focus is often
on extracting reliable inferences from complex, messy datasets:
nonstationary time series, spatial correlations,
missing or biased observations, and the outputs of
computationally intensive climate models. The implied measures
of success are perhaps often pragmatic:
robustness, interpretability, and relevance to the bigger questions. 
This camp exemplifies the translation of statistical rigour
into actionable understanding, perhaps requiring compromises
that a purely theoretical statistician in some cases
might find unsatisfyingly rough. The university methodologists,
on the other side of a perhaps low fence, would on occasion
be more interested in the careful development of models
and accurate optimal statistical analyses (model selection,
estimation, testing, prediction, melding different information
sources, uncertainty quantification),
with clearer recipes for clearer sets of assumptions.

Sometimes the same words refer to rather different concepts in
the two communities. This is particularly true of the term `model', 
which for statisticians are mostly empirical, often low-dimensional
models, while climate science recognises `conceptual models'
of low dimensionality, and `quasi-realistic numerical models'
with very many degrees of freedom;
see~\citet{MuellerStorch04}.

To set the scene for my chapter
I start with a climate-related time series
dataset in Section~\ref{section:bjoernholt}, where
what is being measured is the number of skiing days at
a certain place close to Oslo. Will we be going skiing,
a dozens years from now? Tools involve not only traditional
time series modelling, with a basic autocorrelation
from winter to winter (and there is a fifteen-year gap
in the series), but also certain extensions and variations,
with consequences also for the accuracy of predictions.
These themes are present also in Section \ref{section:future},
working with long monthly time series of temperature anomalies,
from the the National Centers for Environmental Information
database, going back in time to about 1850 and then up to the present.
Sometimes the use of segmented regression,
or broken-line trend curves, is fruitful,
with appropriate fine-tuning, as demonstrated there. 
For some purposes I advocate the use of {\it confidence distributions},
to appreciate and to communicate both basic estimates
and the accompanying statistical uncertainty.
Attempting to predict when in our future the mean trend
of some temperature time series will cross the barrier of
1.5$^{\circ}$C above the average level across 1900--2000
(inside a certain relevant scenario), the point estimate
might be the year 2056, but the relevant statistical distribution
is typically very skewed, with consequences for confidence intervals.
For one of the time series considered there, the point
estimate is indeed the year 2056, but the 90 percent confidence
interval for that unknown year would be from 2026 to infinity -- i.e.,
there is a positive chance that the barrier will never be crossed. 

Climate statistics is of course important also for its
many and partly dramatic implied consequences,
from demography and economy to sociology and biology.
In Section \ref{section:HjortKola} a certain time series
pertaining to the quality of the North Arctic cod,
the skrei, is worked with, ostensibly the longest
teleost time series of marine science, stretching
back to 1859. The point, for the present purposes,
is to analyse that series as potentially influenced by Kola
monthly temperatures, for which there is a hundred years' worth
of time series. In Section \ref{section:iiccff}
themes related to the combination of different sources
of information are discussed. Methodology briefly discussed
there ought to have promise for e.g.~combining predictions
from different climatic models. 
Then in Section \ref{section:Bolt} methods for both
computing probabilities for extreme events and for assessing
the associated uncertainty are illustrated. Again,
confidence curves are revealing the drastic skewness
of the relevant uncertainty distributions;
we might estimate that a certain drastic event next year 
has probability 3.5 percent, but the 90 percent confidence
interval could be the full range from 0 to 19 percent.
A little list of yet further themes of relevance,
for the potential useful of more collaboration
between climate scholars and methodology statisticians,
is provided in Section \ref{section:concluding}.


\section{Where are the snows of yesteryear} 
\label{section:bjoernholt}

\begin{figure}[h]
\centering
\includegraphics[scale=0.35]{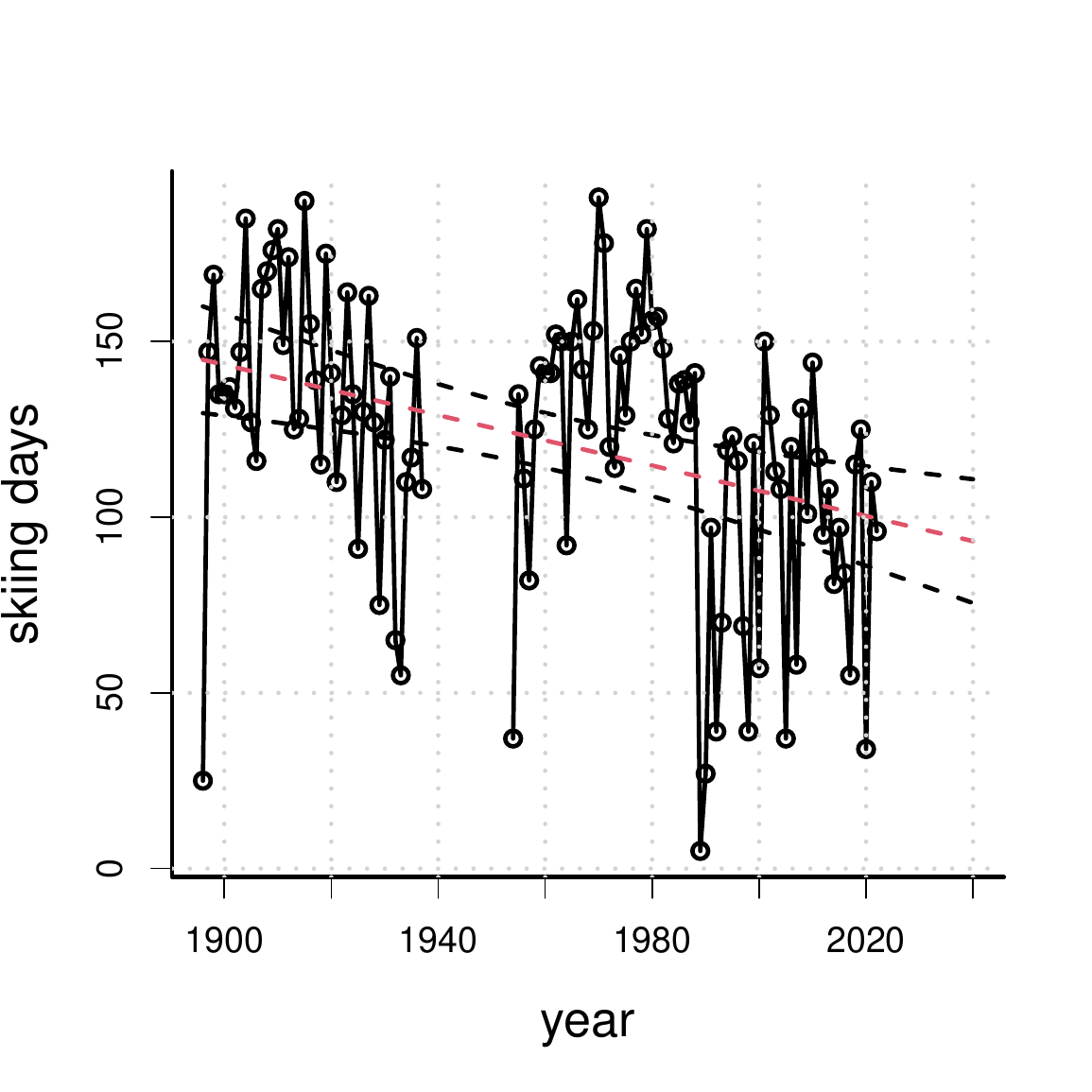} 
\includegraphics[scale=0.35]{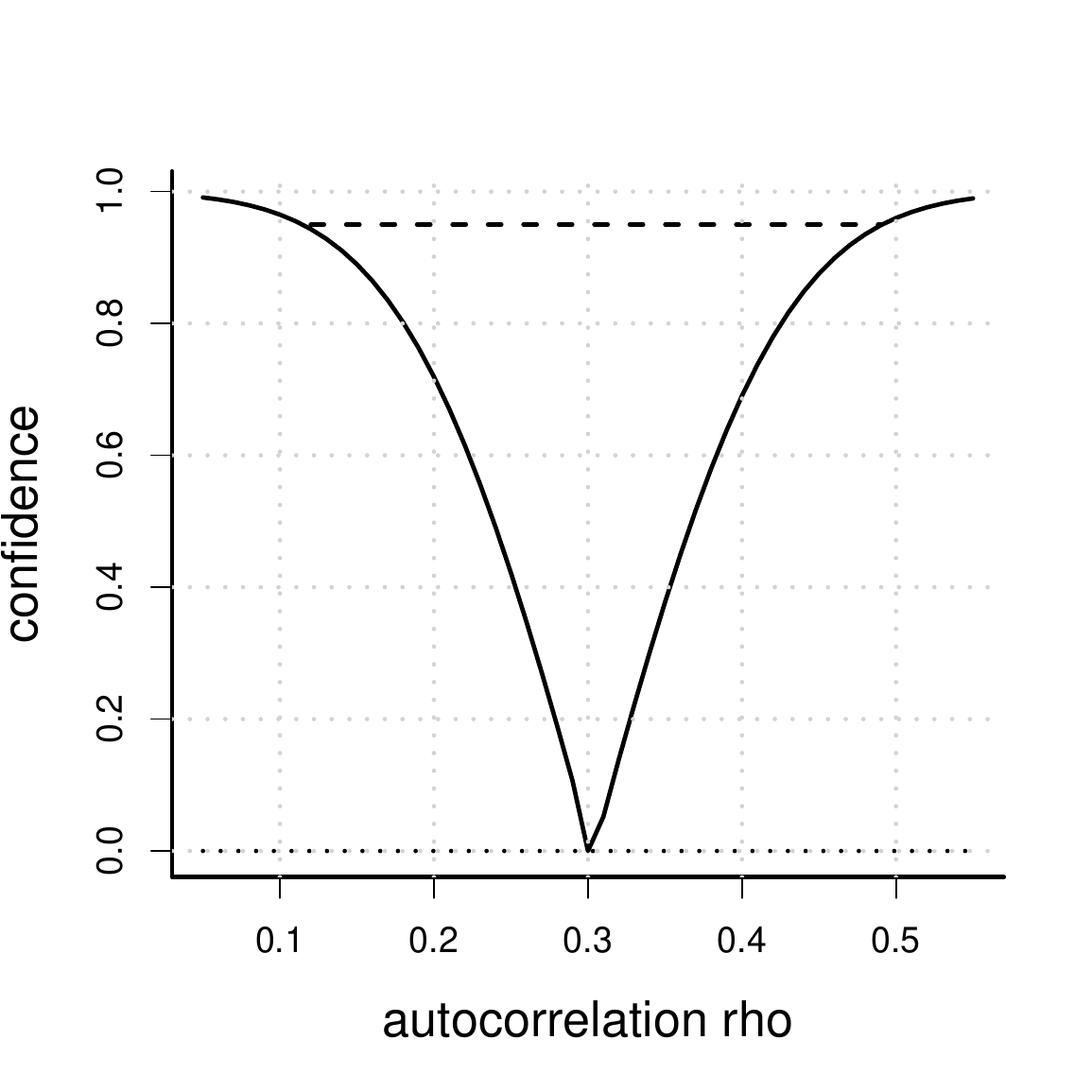} 
\caption{\sl 
Left panel: 
the number of skiing days per year, at the location 
Bj\o rnholt close to Oslo, from 1896 to 2022, though
with a gap in the series, with no records from 1938 to 1954. 
The dashed line is the estimated regression from 
the four-parameter autoregressive model, with 90 percent
confidence band. 
Right panel:
The confidence curve $\cc(\rho)$ for the autocorrelation
parameter of the residuals $y-a-bx$,
with point estimate $0.303$ and 95 percent interval $[0.116,0.490]$.}
\label{figure:skiing_figure11}
\end{figure}

Quo vaditis, Norwegians?
Figure \ref{figure:skiing_figure11} (left panel)
is a potentially dramatic one, for core segments
of the Norwegian population, displaying the number of skiing days per year, 
from 1896 to 2022, at the location Bj\o rnholt in Nordmarka,
a tram distance and a skiing hour north of central Oslo.  
A skiing day is defined as there being at least 25 cm snow
on the ground. How clear is the downward trend, will we still
be able to go skiing, a dozen years from now? 

Classical linear regression remains a powerful tool,
also when dealing with data of this type,
though often enough with extra caveats and modifications,
as I briefly illustrate here. The standard setup
is that of $y_t=a+bx_t+\sigma\eps_t$, with $x_t$
here taken as calendar year minus 1900, and the $\eps_t$
being independent standard normals. This gives
the estimated regression line $144.85 - 0.37\,x_t$
with a fairly large $\hatt\sigma=36.72$. The residuals
carry a clear autocorrelation, however, so a better model
takes $\cov(y_t,y_u)=\sigma^2\,\rho^{|u-t|}$;
see relevant discussion in \citet[Ch.~10]{Storch02}. 
There is a gap in the data, from 1938 to 1953,
so analysis is not very straightforward, but can
be handled. 
  This four-parameter model can be succinctly given as
  $y\sim\N_n(X\beta,\sigma^2 A_\rho)$, for the $n=111$ years
  for which there are observations, with the $a+bx_t$
  structure for the mean and the appropriate correlation
  matrix $A_\rho$ of size $n\times n$. Maximising the log-likelihood
\beqn
\ell=-n\log\sigma-\half\log|A_\rho|
   -\half(y-X\beta)^\tr A_\rho^{-1}(y-X\beta)/\sigma^2
   -\half n\log(2\pi) 
\eeqn 
leads to almost the same regression line as above,
$143.38-0.36\,x_t$, with $\hatt\sigma=36.70$ about
as earlier, but with the crucial autocorrelation being
present, estimated at $\hatt\rho=0.30$.
See Figure \ref{figure:skiing_figure11} (right panel)
for the confidence curve $\cc(\rho)$.
Sadly, the $\hatt b=-0.36$ is very significant,
with 3-4 more lost days for each ten years. 
The autocorrelation does not influence the regression line
much, for this dataset, but has consequences for
prediction and the associated prediction intervals.

We pause for a minute to briefly discuss
{\it confidence distributions} (CDs) and {\it confidence curves} (ccs).
A CD is a random cumulative distribution function for the
parameter under focus, say $C(\gamma,\data)$, with the property that
$\Pr_\gamma(C(\gamma,\data)\le u)=u$ for all $u$ in the unit interval.
In that equation the $\data$ are random, sampled from
the model in question, under the fixed $\gamma$. 
So by inverting $C(\gamma,\data)=0.05$ and $C(\gamma,\data)=0.95$,
for example, for the given dataset, say $\data_\obs$, 
we have a 90 percent confidence interval. From a CD we may form
the convenient and informative confidence curve
\beqn
\cc(\gamma,\data_\obs)=|1-2\,C(\gamma,\data_\obs)|, 
\eeqn
with Figure \ref{figure:skiing_figure11} (right panel) being
one such, and with a few others to follow in sections below.
We may read off a 90 percent interval for the parameter
of interest via $\{\gamma\colon\cc(\gamma,\data)\le0.90\}$, etc.
In this brief explanatory paragraph we have found $\cc(\gamma,\data)$
via first having built $C(\gamma,\data)$, but there
are various situations and recipes for building he confidence
curve directly; see \citet{XieSingh13, SchwederHjort16}
for broader discussion. The CDs and the ccs are reporting
tools to convey both the crucial point estimates and the
associated uncertainty, clearly beyond the $\pm1.96\,\sd$
format under approximate normality. Many estimators have
skewed distributions, leading to skewed confidence intervals,
as then conveyed by the CDs and the ccs. 

Two extensions or modifications of the above, always
worth looking into (though without real need to do so
for the skiing days data, as it turns out) are as follows.
(i) First, the variability may change over time, which may
be tested for and assessed in different ways.
An effective modelling strategy is to work with
$\sigma_t=\sigma\exp(\gamma_1 w_t+\gamma_2 w_t^2)$,
writing $w_t=(x_t-\bar x)/\sd(x)$ for the normalised
$x_t$; this helps numerics and interpretation, with
the $\sigma$ now being the standard deviation for
the middle part of the data. The 6-parametric log-likelihood
can be set up and maximised, etc.; on this occasion
the $\gamma_j$ parameters are small and insignificant, however.
(ii) Second, such data may exhibit fatter tails than those
implied by normality. One may extend the usual model via
$y_t=a+bx_t+\sigma\eps_t$, now with the $\eps_t$
coming from a $t_\nu$ distribution. Again, log-likelihood
functions may be worked out and maximised,
and on this occasion there is no real gain in using t
distributions.

The apparent negative jump for the year 1993 is intriguing,
with no clear explanation. The dataset above is primarily
intended as an illustration of general statistical
techniques, and an attempt at deeper climatic analysis
would have to involve similar time series for other locations,
along with a spatio-temporal apparatus. From the data alone,
the 1993 jump is within behaviour explained by the model,
partly due to the moderately high standard deviation.
There could be other reasons, related to the exact local
conditions where measurements are made, or inhomogeneity,
or to a higher-order autoregressive memory for the residuals,
than for the AR(1) used here. 

When comparing candidate models for the same time series data,
as here, a standard tool is the AIC,
the Akaike Information Criterion, defined as
\beq
\aic=2\,\ell_{\max}-2\dim,
\label{eq:aic} 
\eeq 
twice the maximised log-likelihood minus twice the parameter
dimension. For the skiing days data, the best model
remains the linear trend with autocorrelation,
dealt with above. For certain more specialised purposes,
like optimal prediction for the coming ten years, say,
versions of the Focused Information Criterion
\citet{ClaeskensHjort03, ClaeskensHjort08}
may instead be used, ranking candidate models
by their estimated performance for the given
specialised task. See Section \ref{section:concluding},
Remark B. 




\section{When will we reach 
  1.5\textbf{\boldmath$^\circ$}C above 1900--2000 mean level?}
\label{section:future}

The National Centers for Environmental Information website
provides long temperature time series, at almost any location
on our planet, from 1850 to the present. Here I use such data
at position 60 degrees longitude and 10 degrees latitude,
i.e.~more or less Oslo. The data are given as `temperature anomalies',
year by year, `land and ocean', for each of the twelve months.
These anomalies are the max or the min, inside the months
in question, with respect to the 1991--2020 average.
Below I attempt to exhibit both accurate modelling,
going beyond the more familiar direct linear regression
models pointed to in Section \ref{section:bjoernholt},
and to develop machinery for not only estimating
the future year at which a new barrier is broken,
but also the ensuing and sometimes drastically skewed
uncertainty associated with such estimates. 

\begin{figure}[h]
\centering
\includegraphics[scale=0.35]{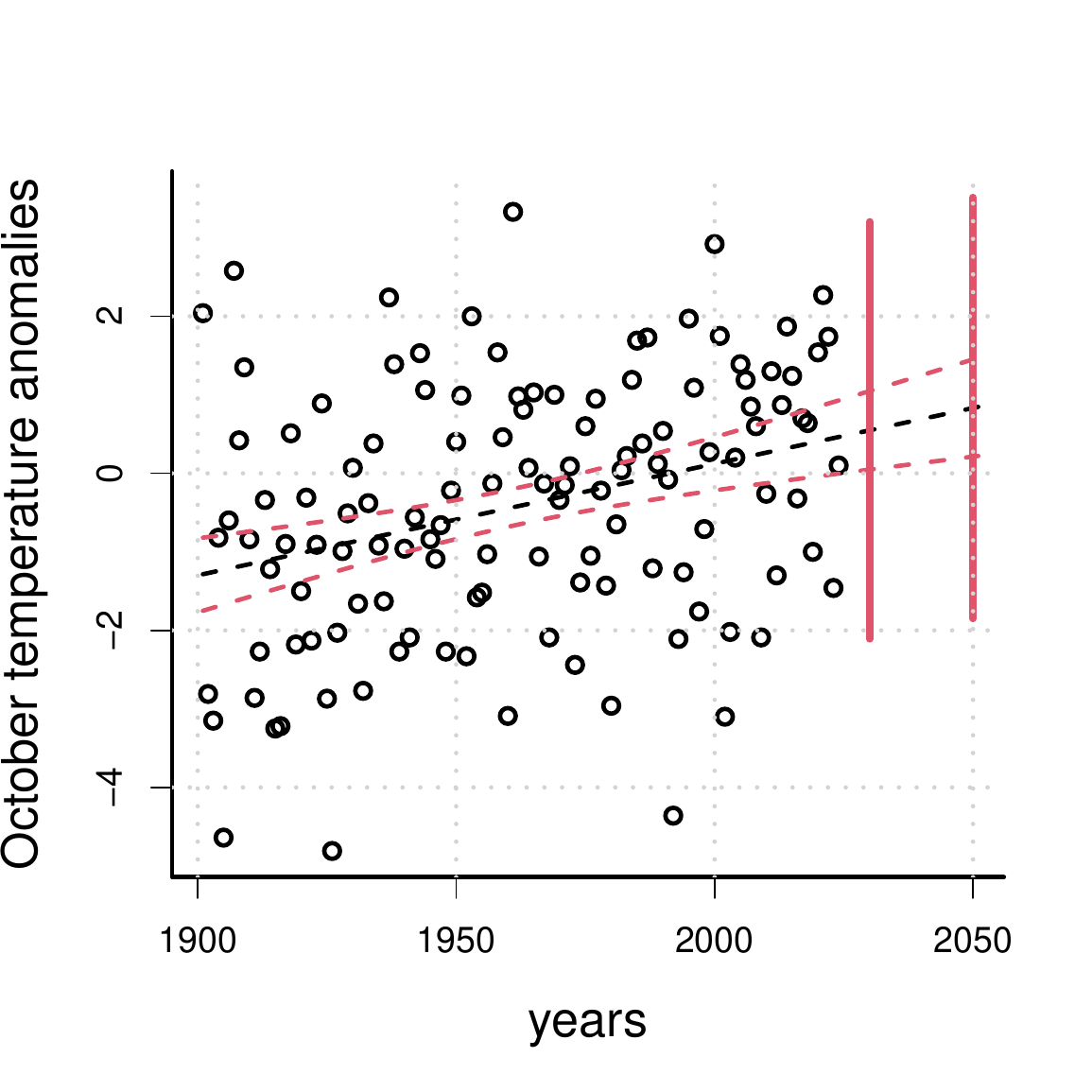} 
\includegraphics[scale=0.35]{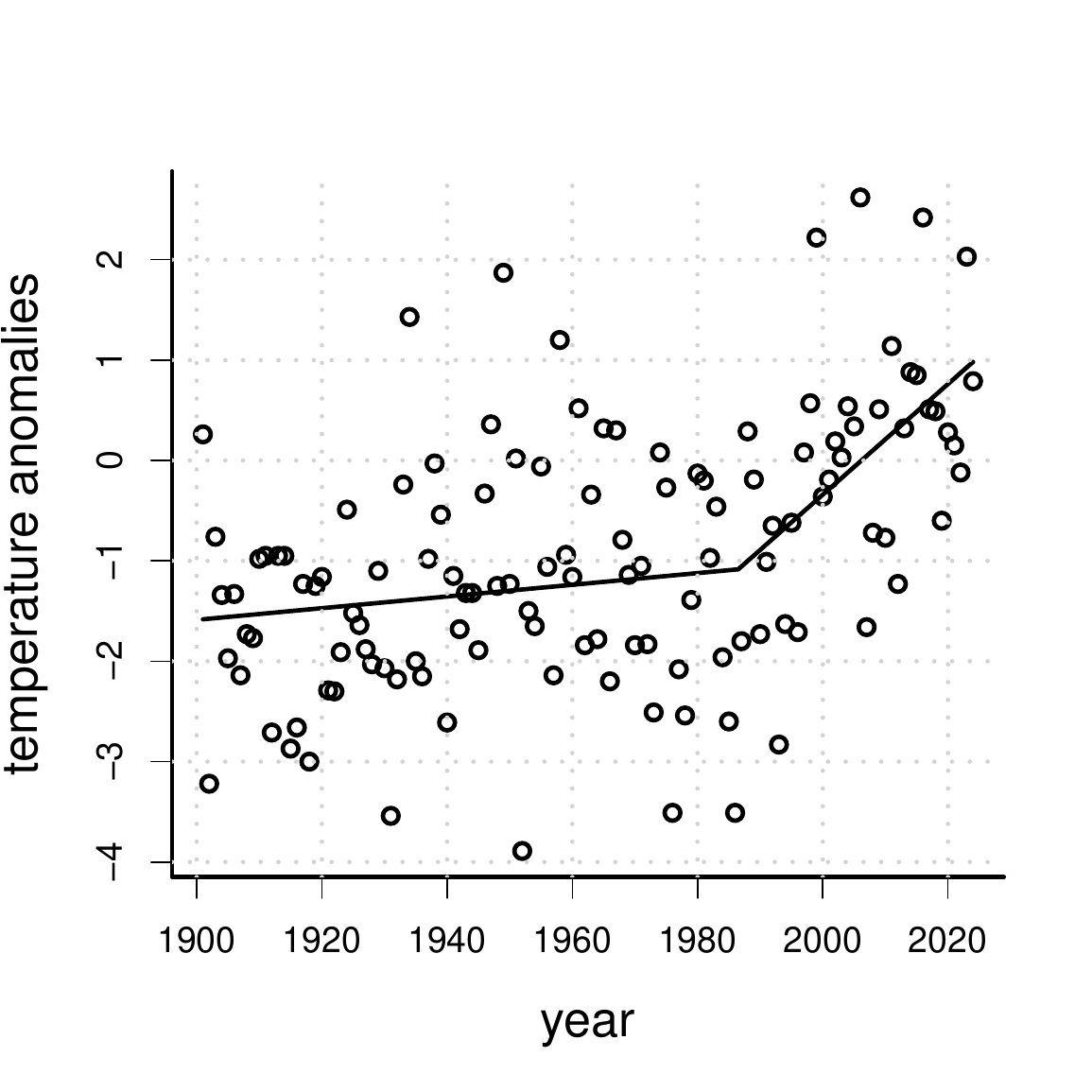} 
\caption{\sl 
  Average monthly temperature anomalies, land and sea,
  at Oslo position, from 1901 to 2024.
  Left panel: for month October, with pointwise 90 percent
  confidence interval for the regression line, along
  with 90 percent prediction intervals for the values
  of $y$ themselves, for 2030 and 2050.
  Right panel: for month September, using 
  connected segmented regression. The estimated break-point,
  from small derivative to strong derivative, 
  is at 1986.}
\label{figure:hamburg51}
\end{figure}

Considering the October series,
the linear regression model $y_t=a+b(x_t-\bar x)+\sigma\eps_t$
is adequate,
for data pairs $t=1,\ldots,n$, with the $\eps_t$ standard normal;
also, $\bar x$ is the average over these years. 
Figure \ref{figure:hamburg51} (left panel) displays the data,
along with the fitted regression line and a pointwise
90 percent confidence band, stretched all the way up to 2050.
Crucially, this relates to predicting the line itself,
as opposed to predicting the October temperature
anomalies themselves for these years. Clearly, we
can predict the October 2050 outcome with far less
precision than for its expected value. An effective tool there,
to predict $y_\new$ at year $x_\new$, is
the confidence distribution (CD) 
\beq
C(y_\new,\data)=G_m\Bigl( {y_\new-\hatt a-\hatt b(x_\new-\bar x)
     \over \hatt\sigma\{1+1/n+(x_\new-\bar x)^2/M_n\}^{1/2}}\Bigr), 
\label{eq:CDone}
\eeq 
in terms of the c.d.f.~of the $t_m$ distribution, with $m=n-2$
degrees of freedom; also, $M=\sumin(x_i-\bar x)^2$.
This is used in the figure to plot 90 percent confidence intervals
for the October $y_\new$ at 2030 and 2050. 

\begin{figure}[h]
\centering
\includegraphics[scale=0.35]{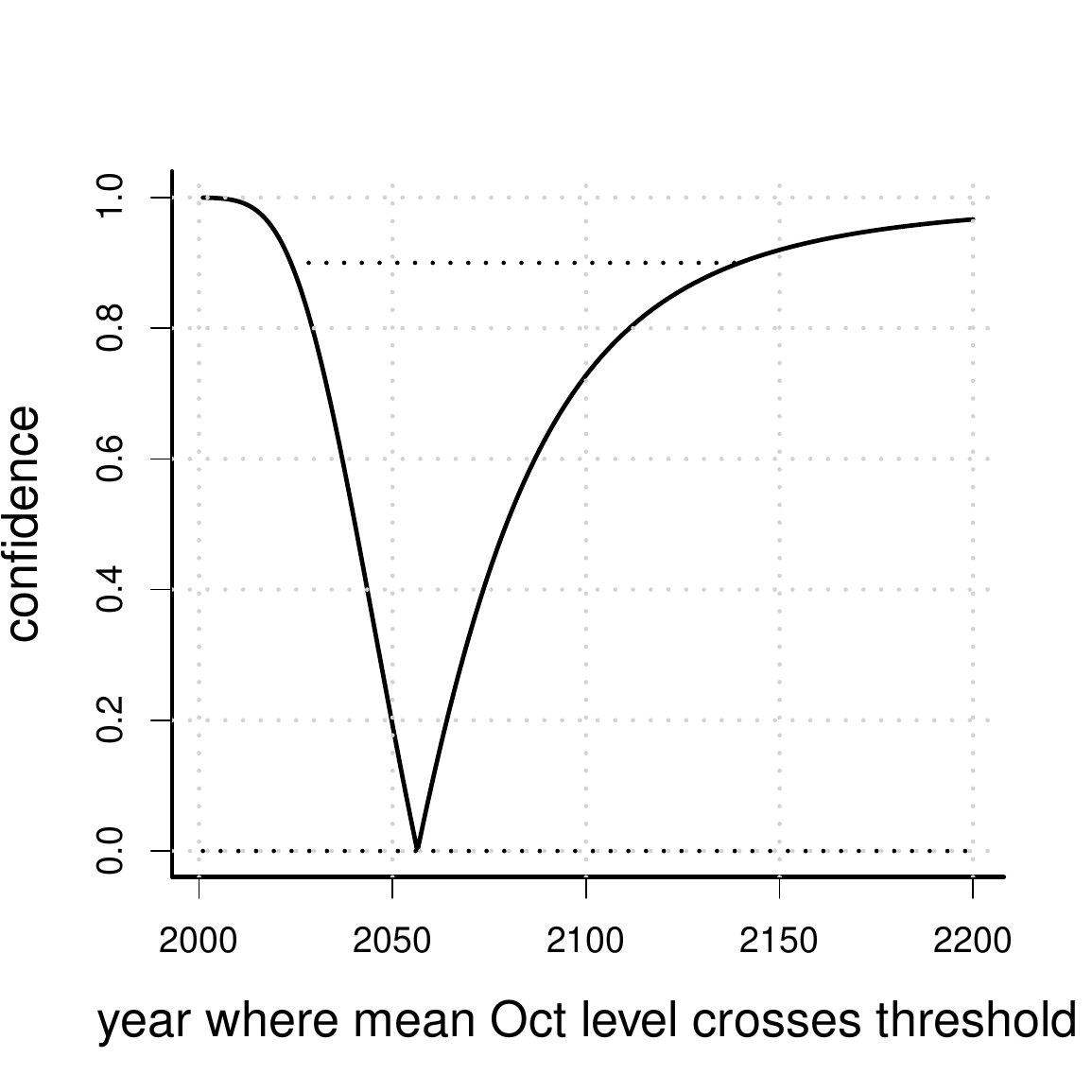}
\includegraphics[scale=0.35]{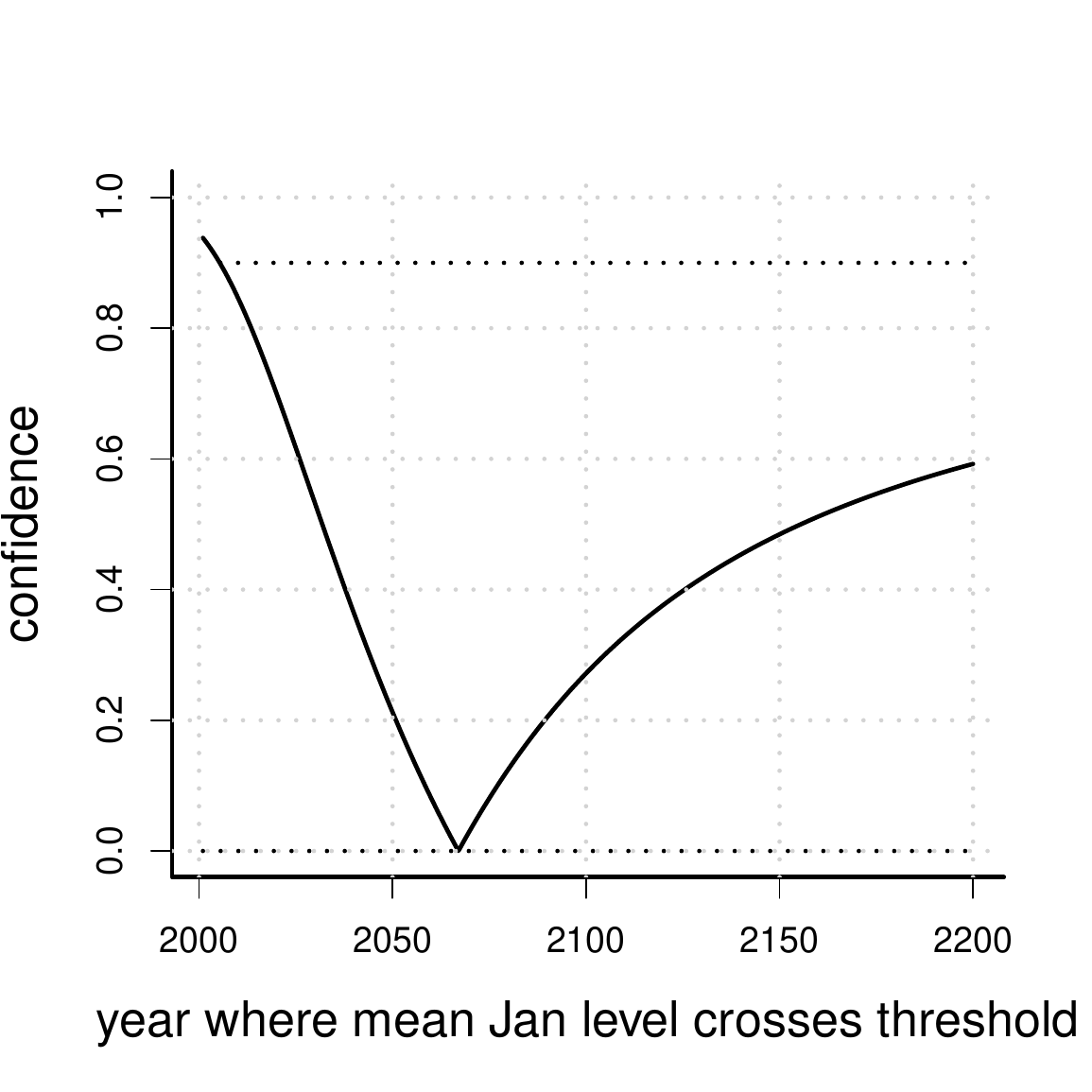}
\caption{\sl
  At which year $x_0$ in the future will the mean level
  $a+b(x_0-\bar x)$ reach the threshold $y_0$, defined as
  1.5$^\circ$ C above the average value over 1901--2000,
  for the month in consideration? 
  Here are confidence curves $\cc(x_0)$,
  for October temperatures (left panel)
  and January temperatures (right panel).
  For October, the point estimate is 2056, with
  90 percent interval from 2024 to 2139, skewed to the right.
  For January, the skewness is stronger, and the confidence
  curve never goes above 0.864; confidence intervals
  at levels higher than this will include infinity,
  i.e.~the threshold will never be reached. The climate
  increase is stronger for the summer months than for the winter months.}
\label{figure:hamburg53}
\end{figure}

As long as the regression derivative $b$ is estimated as positive,
as here, it makes sense to to ask about the future year $x_0$
where the level $a+b(x_0-\bar x)$ reaches some given threshold $y_0$.
This corresponds to $x_0=\bar x+(y_0-a)/b$, with point estimate
$\hatt x_0=\bar x+(y_0-\hatt a)/\hatt b$. Rather than attempting
to use the delta method, which will not work well for
small $b$, we may use the fact that $y_0-\hatt a-\hatt b(x_0-\bar x)$
is zero-mean normal with variance
$\sigma^2\{1/n + (x_0-\bar x)^2/M_n\}$. This leads to the CD
\beq
C(x_0,\data)=F_{1,m}\Bigl( { \{y_0-\hatt a-\hatt b(x_0-\bar x)\}^2
  \over \hatt\sigma^2 \{1/n + (x_0-\bar x)^2/M_n\}^{1/2} }\Bigr), 
\label{eq:CDtwo}
\eeq
with $F_{1,m}$ the c.d.f.~for the F distribution with degrees of
freedom $(1,m)$, with accompanying confidence curve
$\cc(x_0,\data)=|1-2\,C(x_0,\data)|$. 
This machinery is illustrated in Figure \ref{figure:hamburg53},
for the NCEI data, for months October and January.
Note here that as $x_0$ grows, there is a finite limit below 1,
namely $F_{1,m}(M_n\hatt b^2/\hatt\sigma^2)$,
which can be seen to be $1-p^*$, with $p^*$ the p-value of
the natural test for $b=0$ vs.~$b\not=0$.
In situations with moderate p-values,
i.e.~$t_n=M_n^{1/2}\hatt b/\hatt\sigma$ small or moderate,
there is a clear positive chance that the temperature level
will {\it never} reach the level in question.
For this illustration, the $t_n$ values are the moderate 1.501 
and the very clear 3.598 (on the t scale with $m=n-2$ degrees
of freedom), for January and October, reflected in the plots
of Figure \ref{figure:hamburg53},

\begin{figure}[h]
\centering
\includegraphics[scale=0.35]{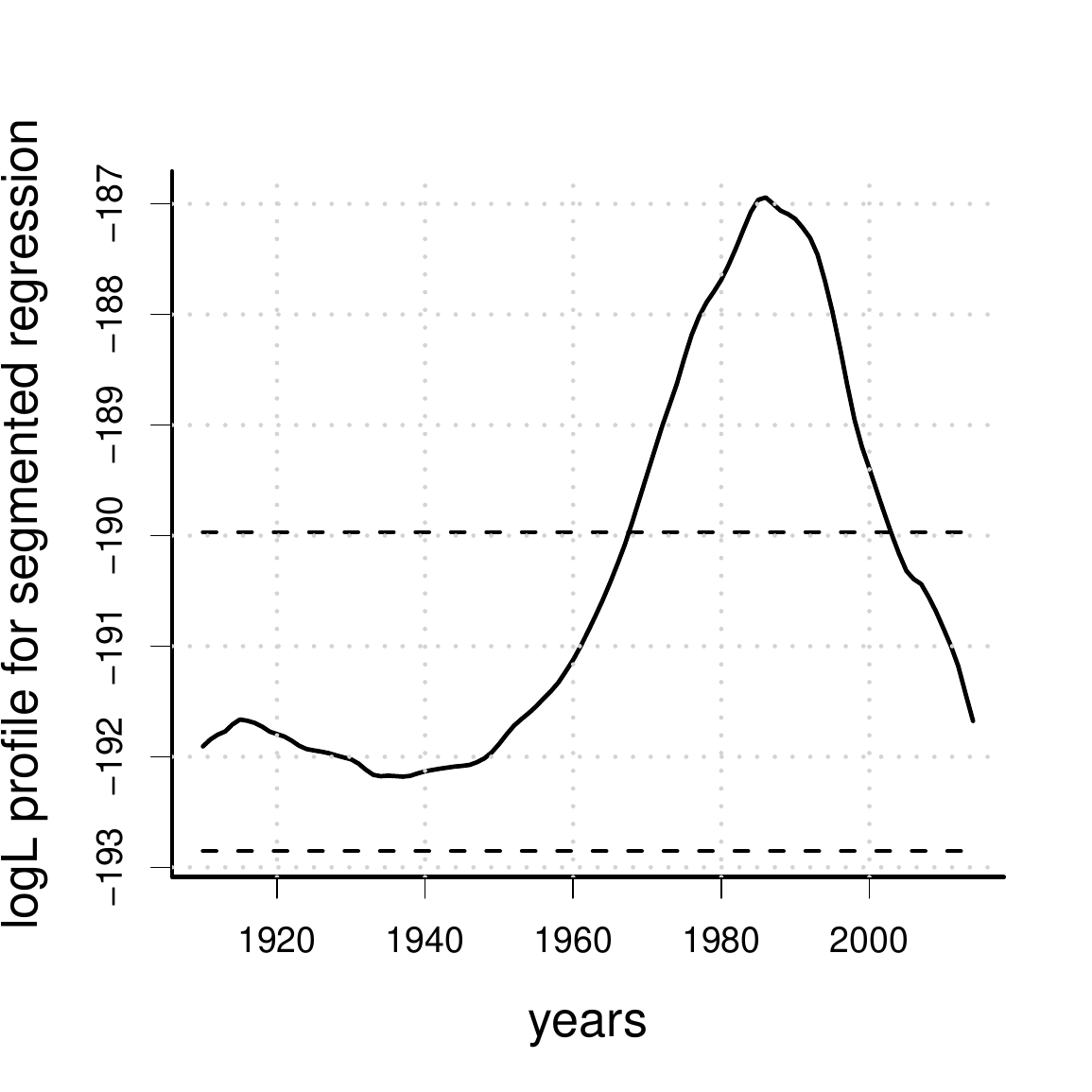} 
\includegraphics[scale=0.35]{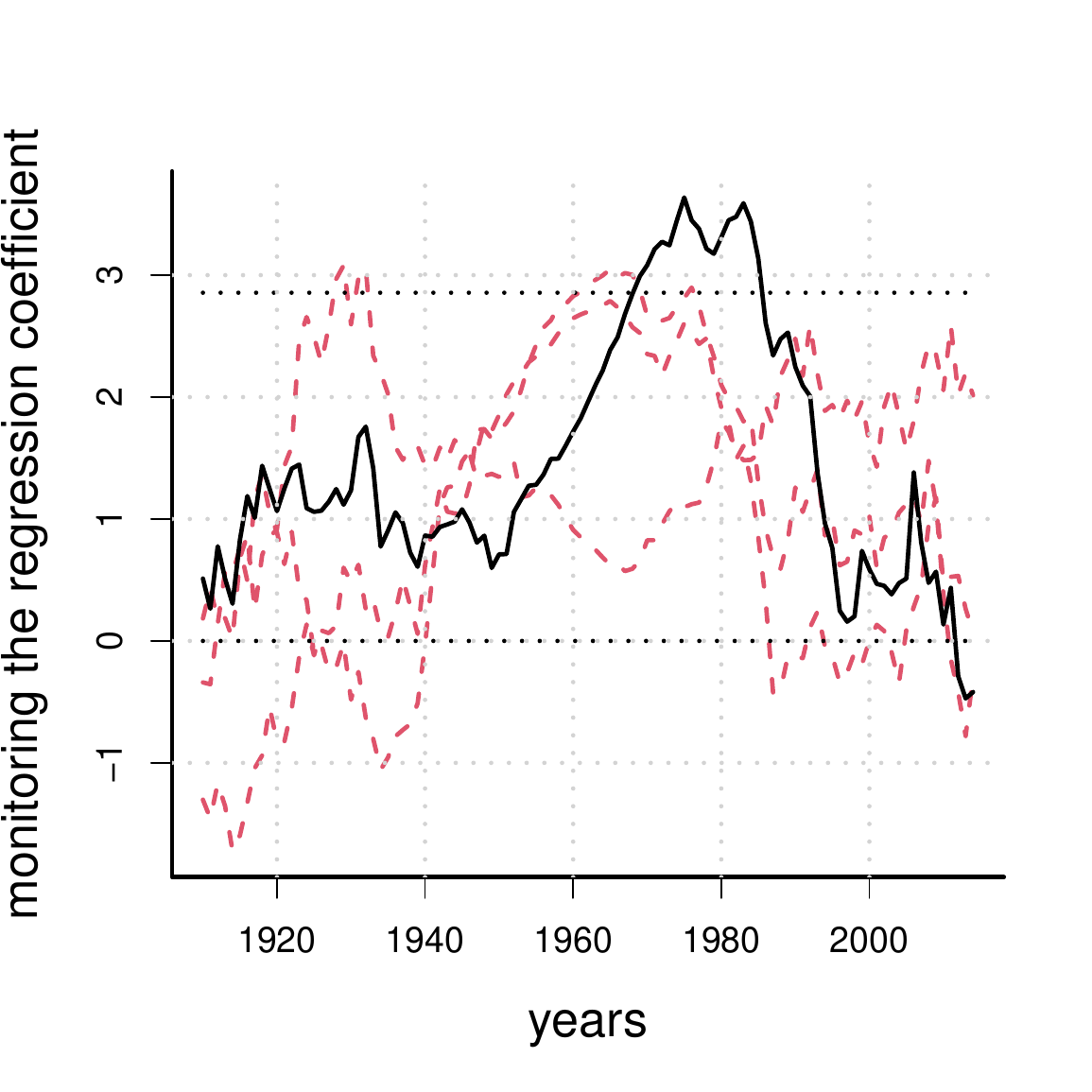} 
\caption{\sl
  Left panel: log-likelihood profile for the segmented
  regression model, for the September series,
  with maximum very significantly exceeding both the
  log-likelihood maxima
  $\ell_{\max,1}$ of linear regression and
  $\ell_{\max,2}$ of quadratic regression. 
  Right panel: monitoring plot $B_n$ for the regression coefficient
  $b$ in $y_i=a+bx_i+\eps_i$, for months January, February,
  June, September; the plot for September reaches
  above the 0.95 null distribution quantile 2.856,
  indicates that $b$ for that month has not remained constant.}
\label{figure:hamburg61}
\end{figure}

Examining the September time series 1901--2024
I shall exhibit so-called segmented regression, useful
also for studying structural breaks. The model takes
\beqn
y_t=\begin{cases} a_L+b_L(x_t-\bar x)+\sigma\eps_t &{\rm for\ } t\le\tau, \\
a_R+b_R(x_t-\bar x)+\sigma\eps_t &{\rm for\ } t>\tau, 
\end{cases} 
\eeqn 
for an appropriate break point $\tau$, inside say $i_0,\ldots,n-i_0$,
with $i_0$ the minimum size for a series to take form.
Here we wish the left and right parts to be knotted together in
a continuous fashion, so we stipulate that
$a_L + b_L (\tau+\half-\bar x)= a_R + b_R(\tau+\half-\bar x)$.  
With $\sigma$ the standard deviation for the i.i.d.~normal $\eps_i$,
the model has hence $3+1+1=5$ parameters. The log-likelihood
profile function in the break point $\tau$ takes the form
\beqn
\ell_\prof(\tau)=\max_{{\rm all\ }a_R,b_L,b_R}\{-n\log\sigma
   -\half Q(a_R,b_L,b_r)/\sigma^2\}, 
\eeqn 
with
\beqn
Q(a_R,b_L,b_R)=\sum_{t=1}^\tau \{y_t-a_L-b_L(x_t-\bar x)\}^2
   +\sum_{t=\tau+1}^n \{y_t-a_R-b_R(x_t-\bar x)\}^2,  
\eeqn 
where we have used $a_L$ solved with respect to $a_R,b_L,b_R$.
Figure \ref{figure:hamburg61} (left panel) displays
this profile function, peaking at the year 1986,
indicating that a weak $b$ at that point has changed
to a bigger $b$; see Figure \ref{figure:hamburg51} (right panel)
for the resulting fitted curve, with the 2001--2024 data.
Note also that the log-likelihood max is significantly above
those for the linear and quadratic regression models,
and that the segmented regression is winning the AIC
ranking associated with (\ref{eq:aic}). The AIC formula
itself stems from assumptions involving models being smooth
in its parameters, so there are mild technical issues
here, but the increase in log-likelihood here, favouring
the segmented model, can be shown to be clearly significant.

We are learning via Figure \ref{figure:hamburg61} (left panel)
that the segmented regression model is a fitting one
for the September series, better than both linear and quadratic
regression. For additional insight, and for checking
with the other months, we may form
a monitoring process to check if the $b$ regression
coefficient for a $y_i=a+b(x_i-\bar x)+\eps_i$ model has remained constant
over time, as follows. For any candidate changepoint $\tau$
for this derivative, compute first the estimated coefficients
$\hatt b_L$ and $\hatt b_R$ for the left part ($1,\dots,\tau$)
and the right part ($\tau+1,\ldots,n$), along with
$M_L(\tau)=\sum_{t=1}^\tau (x_t-\bar x_L)^2$ and
$M_R(\tau)=\sum_{t=\tau+1}^n (x_t-\bar x_R)^2$,
with $\bar x_L$ and $\bar x_R$ the averages.
Then the process with values 
\beq
B_n(\tau)={\hatt b_R-\hatt b_L\over
  \hatt\sigma \{1/M_L(\tau)+1/M_R(\tau)\}^{1/2} }
\quad {\rm for\ }\tau=i_0,\ldots,n-i_0, 
\label{eq:bridgeforb}
\eeq
with $i_0$ the minimum length for forming a meaningful
regression dataset, and with $\hatt\sigma$ the estimated
residual standard deviation, is under no-change conditions
close to a well-defined weighted Brownian bridge.
Formally, with $Z_n(s)=B_n([ns])$, it is demonstrated in
\citet[Ch.~10]{HjortStoltenberg26} that there is full process
convergence of $Z_n$ to $Z(s)=W^0(s)/\{s(1-s)\}^{1/2}$,
over each $[\eps,1-\eps]$ subinterval of the unit interval. 
Such $B_n(\tau)$ are plotted in
Figure \ref{figure:hamburg61} (right panel),
for months January, February, June (dashed curves),
for illustration, not demonstrating any clear change
in the $b$ coefficient, whereas the monitoring plot
for September is clearly significant. 

We may also investigate the temperature anomalies series
1901--2024 for the other months. 
For some of these, the fit is better for the quadratic
regression model $y_t=a+b(x_t-\bar x)+c(x_t-\bar x)^2+\sigma\eps_t$,
as judged by the AIC of (\ref{eq:aic}) and other related
criteria, with $\eps_i$ standard normal,
and with the $\sigma$ now changing interpretation, 
for the residuals with respect to the quadratic fit.
There are natural extensions of the (\ref{eq:CDone})--(\ref{eq:CDtwo})
CD formulae to the case of cubic regressions,
for a future mean level as well as for the barrier-crossing
future year.
Analyses reveal that the climate increase is strongest
for the summer months April to October, 
and rather weaker for the winter months November to March. 

\section{The Hjort liver index and Kola temperatures} 
\label{section:HjortKola}

In climate statistics, one often wishes to connect the statistical
modelling and analyses of climate change to its downstream impacts
on human and ecological systems. Such effects abound,
in domains including human impacts (migration, health, economy,
psychology); terrestrial and other biology (biodiversity, abundance);
and marine biology (ocean warming, acidification, coral bleaching,
species shifts). As an instance inside this broad terrain,
where climate science meets marine biology, I discuss how
{\it Kola temperatures} might have affected the
{\it quality of the skrei}
(the Northeast Arctic codi, Gadus marhua),
where we do have very long time series. 

\begin{figure}[h]
\centering
\includegraphics[scale=0.35,angle=0]{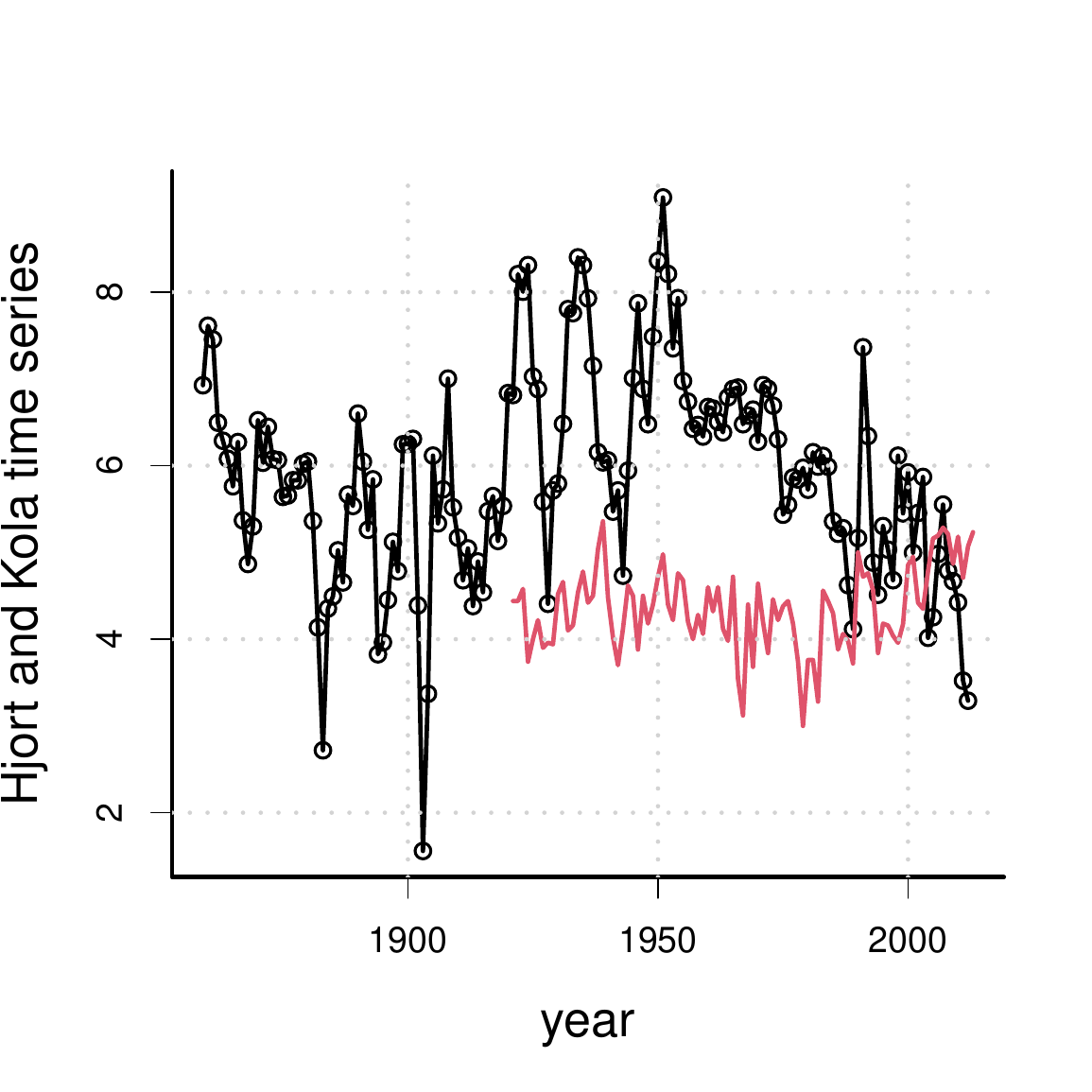} 
\includegraphics[scale=0.35,angle=0]{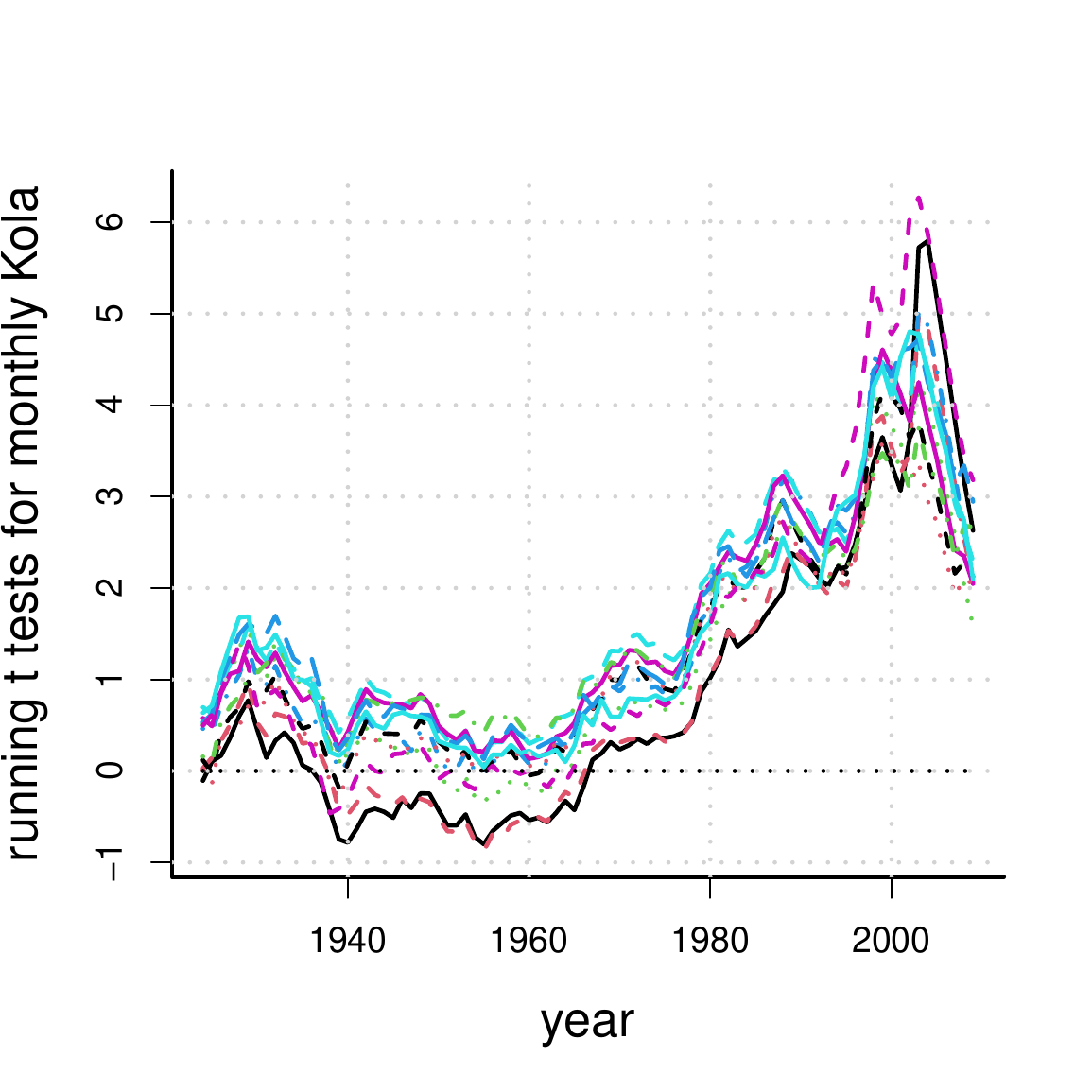} 
\caption{\sl 
Left panel: 
the Hjort liver index, 1859--2013 (percentage of liver 
in the skrei, the Northeastern Atlantic cod) 
with the annual Kola temperature, 1921--2013 (in degrees Celsius).
Right panel: 
running t tests plot for Kola temperatures 1921 to 2013,
one curve for each month.} 
\label{figure:johanhjort41}
\end{figure}

The first four chapters of \citet{JHjort14}, a classic 
in fisheries science and marine biology, essentially pertains 
to the {\it quantity} of fish and the fluctuations of fish populations. 
Hjort was however also concerned with what he terms the 
{\it quality} of fish and devotes most of the book's Chapter 5 
to how this can reasonably be defined, also attempting to 
identify influencing factors. The liver quality index thus 
defined was `no.~of hectolitres of liver per 1000 skrei', 
leading also to one of the first comprehensive teleost
time series ever published, for the time period 1880--1912;
see \citet{Smith94}. 
Later efforts, detailed in \citet*{Kjesbuetal14} and 
\citet*{HermansenHjortKjesbu16}, involving also a more
careful definition of Hjort's Hepatosomatic Index (HSI), 
have led to one of the longest time series in all of fisheries science,
the Hjort Liver Index 1859 to the present. 
Also historically impressive are the data systematically 
collected on monthly Kola temperatures since 1921, 
by Russian marine biologists, summarised in \citet{Boitsovetal12}. 
Figure \ref{figure:johanhjort41} (left panel) shows the HSI series 
along with the annual average Kola temperatures 
(the HSI in percent, the temperatures in Celcius).
Below I go into both the increased Kola temperatures
and how they be seen as influencing the HSI series. 

We start with the Kola temperatures, of clear separate interest. 
To assess whether the apparent increase is significant,
from Figure \ref{figure:johanhjort41} (left panel), 
we study the twelve temperature series, the January temperatures 
up to the December temperatures, from 1921 to 2013. 
For any of these, with temperatures $x_1,\ldots,x_n$, 
we compute first the overall mean $\bar x$ 
and standard deviation $\hatt\sigma$, and also the autoregressive 
first order coefficient, 
i.e.~$\hatt\rho=(1/n)\sum_{t=1}^{n-1} \hatt\eps_{t-1}\hatt\eps_t$,
with $\hatt\eps_t=(x_t-\bar x)/\hatt\sigma$. These turn out
to be close, month for month; their average is $\hatt\rho=0.422$.
For autoregressive means we have in general that
$\Var\,\bar x\doteq(\sigma^2/n)f^2$, with the
extra factor $f^2=(1+\rho)/(1-\rho)$, here
leading to estimated factor $\hatt f=1.567$. 
Then, month for month, we compute running t tests,
taking the AR(1) factor into account, of the type
\beq
T_n(\tau)={\bar x_R-\bar x_L\over
  \{\hatt\sigma_L^2/n_L + \hatt\sigma_R^2/n_R\}^{1/2} \hatt f}
\quad {\rm for\ } \tau=i_0,\ldots,n-i_0, 
\label{eq:bridgeforx} 
\eeq 
with the data split into left $(1,\ldots,\tau)$
and right $(\tau+1,\ldots,n)$ parts.
This leads to Figure \ref{figure:johanhjort41} (right panel).
These monitoring bridges are similar to
those for checking constancy of regression coefficients
in (\ref{eq:bridgeforb}).  
As for that earlier case, there is process convergence of 
$V_n(s)=T_n([ns])$ to $V(s)=W^0(s)/\{s(1-s)\}^{1/2}$,
the naturally weighted Brownian bridge on the unit interval,
under the null change hypothesis.
The distribution of $V_{\max}=\max_{\eps\le s\le1-\eps}|V(s)|$
can be simulated, for any $[\eps,1-\eps]$ window,
with e.g.~upper 5 percent quantile 3.18 for the $[0.025,0.975]$
case. In view of this, the figure very convincingly shows 
that the Kola temperatures have been rising, at least since 1990. 

\begin{figure}[h]
\centering
\includegraphics[scale=0.35,angle=0]{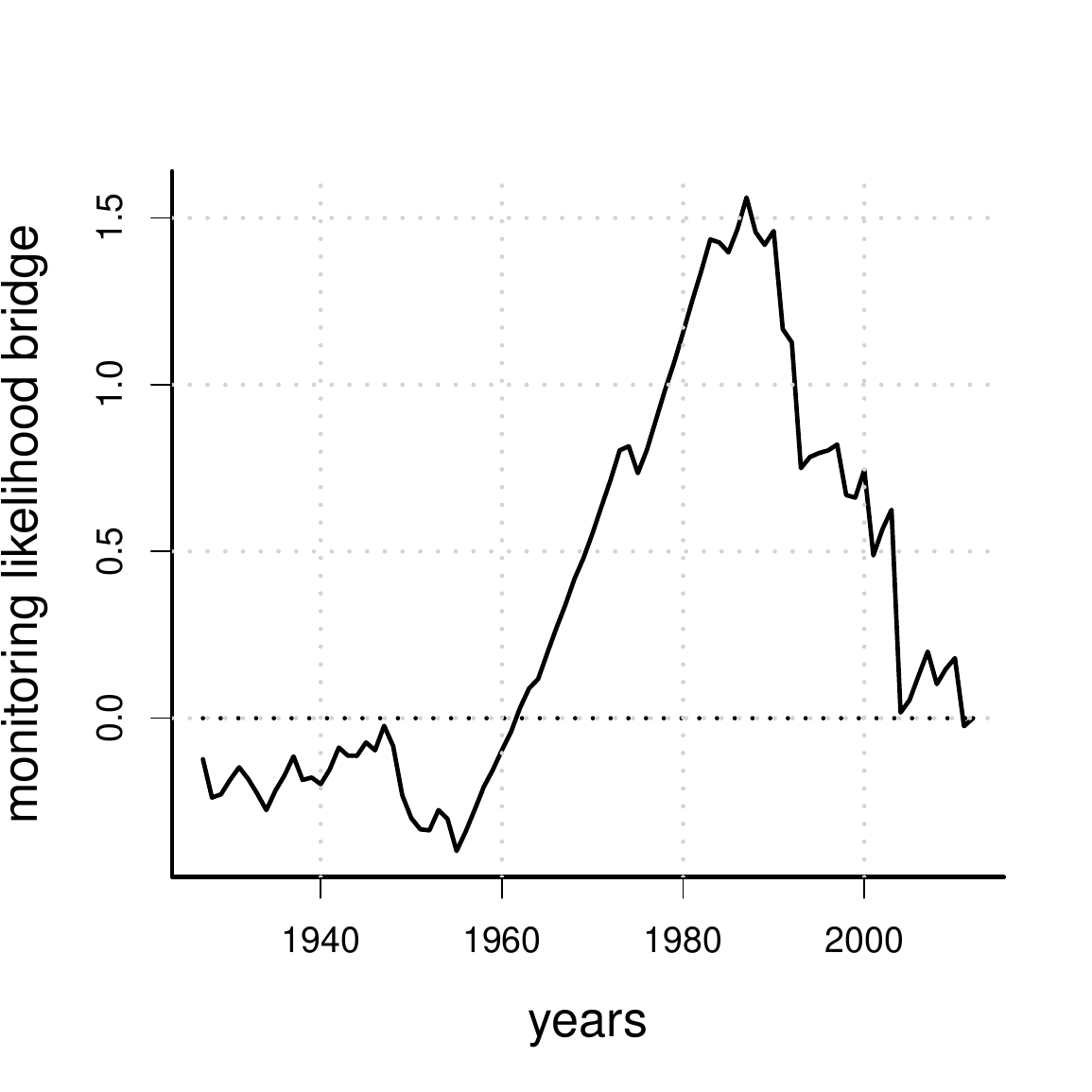} 
\includegraphics[scale=0.35,angle=0]{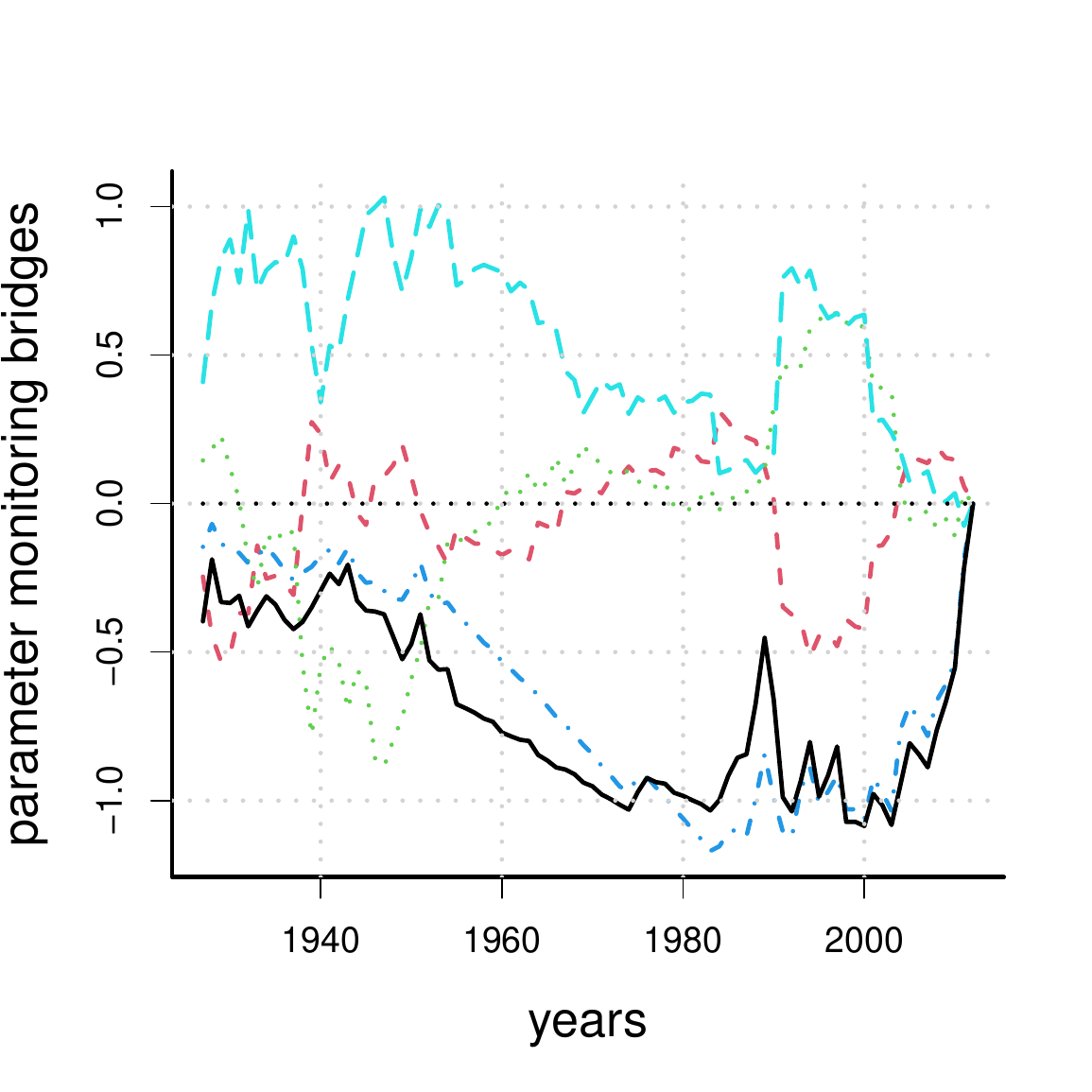} 
\caption{\sl 
Monitoring bridges for testing whether the five-parameter 
model $y_i=\beta_0+\beta_1x_{i-1}+\beta_2x_{i-2}+\sigma\eps_i$,
with the $\eps_i$ having autocorrelation $\rho$, has remained
unchanged. Left panel: log-likelihood maxima bridge,
peaking at 1987, indicating a change, but not a major one. 
Right panel: bridges for each of the five parameters,
indicating that these have remained reasonably constant over time.}
\label{figure:johanhjort53}
\end{figure}

We now come to analysing the influence of the Kola temperatures
on the Hjort index series. 
Taking the skrei's spawning seasons and behaviour into account, 
it is argued in \citet*{HermansenHjortKjesbu16} that the most 
relevant information from the Kola temperatures, when it 
comes to the skrei and its liver quality $y_t$ for year $t$,
is in terms of $x_{t-1},x_{t-2}$, say, 
denoting temperatures from previous winters. 
Specifically, for year $t$, let $x_{t-1}$ be the average 
temperature taken over the five months
October, November, December the previous year
and January, February for the present year.   
These $x_{t-1},x_{t-2}$ may then be computed from the data, 
for the $n=93$ years from 1921 to 2013.
The model
\beq
y_t=\beta_0+\beta_1x_{t-1}+\beta_2x_{t-2}+\sigma\eps_t
\quad {\rm for\ }t=1,\ldots,n 
\label{eq:hjortmodel} 
\eeq 
is now a pertinent and revealing one, 
again with the $\eps_t$ being a unit variance AR(1) series. 
Maximum likelihood theory, valid also for such
time series with covariates, can be applied
to find parameter estimates, standard errors,
and their Wald ratios. For the regression coefficients, 
these are $(0.121,0.168,0.718)$ for $\beta_1$ but a crucially
more prominent $(0.363,0.168,2.154)$ for $\beta_2$.
Thus the liver index for year $t$ is significantly
and positively influenced by the Kola temperature
two years before. There is also strong autocorrelation
here, however, estimated at 0.848 with standard error 0.092,
indicating that the Kola climate environment is crucial
also for years one and three steps back in time. 

Has the model (\ref{eq:hjortmodel}), for how marine temperatures
influence this particular aspect of marine life,
stayed about the same, for these hundred years?
Such statistical questions, also for many other setups
of relevance when studying climatic impact on biology,
can be handled. One relevant method, in this context
of mechanisms developing over time, is to start with
the log-likelihood maxima sequence, and then building
a monitoring bridge for these, as follows. Having observations
for time points $1,\ldots,\tau$, leading to responses
$y_\tau$ of length $\tau$ and covariance matrix $X_\tau$
of size $\tau\times3$, we work with the log-likelihood function
\beqn
\ell_\tau(\beta_0,\beta_1,\beta_2,\sigma,\rho)
=-\tau\log\sigma-\half\log|A_{\tau,\rho}|-\half Q_\tau/\sigma^2
   -\half\tau\log(2\pi), 
\eeqn 
in which $A_{\tau,\rho}$ is the autocorrelation matrix
of size $\tau\times\tau$ and 
$Q_\tau=(y_\tau-X_\tau\beta)^\tr A_\rho^{-1}(y_\tau-X_\tau\beta)$.
Optimisation leads to the maximum $\ell_{\max,\tau}$,
from which we construct the monitoring process
\beqn
Z_n(\tau)=(1/\rootn)\{\ell_{\max,\tau}-(\tau/n)\ell_{\max,n}\}/\kappa
\quad {\rm for\ }\tau=i_0,\ldots,n, 
\eeqn 
with $i_0$ the minimum length for having established
estimates for the five parameters of the model.
The $\kappa$ factor is there to make $Z_n(\tau)$
come close to a Brownian bridge for growing length $n$
of the time series, and can for this model be shown
to be $(1/2)^{1/2}$; see \citet*{CunenHermansenHjort18}. 
The $Z_n(\tau)$ process is displayed in Figure \ref{figure:johanhjort53}
(left panel), peaking at 1987, with maximum value 1.56.
This corresponds to a p-value of 0.02, via the null distribution
under no-change conditions, close to the maximum absolute value
of a Brownian bridge.

This indicates that the five parameters of the (\ref{eq:hjortmodel})
have not remained entirely constant, over the past hundred
years. Any such changes have not been drastic, however,
as we also learn via monitoring bridges for the five
individuals parameters,
shown in Figure \ref{figure:johanhjort53} (right panel).
These are of the form
\beqn
M_{n,j}(\tau) = (1/\rootn)
   \tau(\hatt\theta_{j,\tau}-\hatt\theta_{j,n})/\hatt\kappa_j
   \quad {\rm for\ }\tau=i_0,\ldots,n, 
\eeqn 
for parameters $1,\ldots,5$,
Here $\hatt\theta_{j,\tau}$ is the maximum likelihood
estimates computed after $\tau$ observations, so
the $M_{n,j}$ start and end at zero. The $\hatt\kappa_j$
is the factor making these processes tend to Brownian
bridges for increasing data volume. For the required
details, see \citet[Section 2]{HjortKoning02}.
For the present case, the parameter monitoring processes
stay within the natural range of Brownian bridges
(the upper 0.05 quantile point of the distribution
of its maximum absolute value is 1.358), so there is no
clear indication that any of the parameters have undergone
any drastic changes, across these hundred years
of climatic biology history. The key mechanisms
involved, when understanding how Kola temperatures
influence the Hjort index, have been essentially the same. 



\section{Combination of (perhaps very) different information sources}
\label{section:iiccff} 

Suppose different information sources inform us
about a certain quantity. Combining such is partly standard
terrain in statistical meta-analysis, where the classical 
question considered is that of combining independent estimates
of a common mean, perhaps along with an additional variance
assessment. Often enough information sources are however
much more diverse, and the perhaps multiple quantities
considered more complicated.

In \citet{CunenHjort21} a quite general setup is worked with. 
In brief, their II-CC-FF paradigm for fusing separate
confidence distributions into a combined on runs as follows.
I am satisfied here to showcase the method for
the special case where there is a single focus parameter
$\phi$ to be assessed; more general versions are worked out
and applied in the article mentioned. 
In the Independent Inspection step, separate CDs are built
for the parameter $\phi$ of main interest, say $C_j(\phi)$
for data sources $j=1,\ldots,k$. These lead in turn 
to confidence curves $\cc_j(\phi)=|1-2\,C_j(\phi)|$. 
In the Confidence Conversion step, these are converted
to exact or approximate log-likelihood contributions via
$\ell_j^*(\phi)=-\half G_1^{-1}(\cc_j(\phi))$, which we call
normal conversion, in terms of the quantile function of the $\chi^2_1$.
For the final Focused Fusion step, log-likelihood contributions
are combined to $\ell^*(\phi)=\sumjk \ell_j^*(\phi)$,
which via the deviance $D(\phi)=2\{\ell^*_{\max}-\ell^*(\phi)\}$
and the Wilks theorem yields the fused confidence curve
$\cc^*(\phi)=G_1(\cc^*(\phi))$. 

\begin{figure}[h]
\centering
\includegraphics[scale=0.35,angle=0]{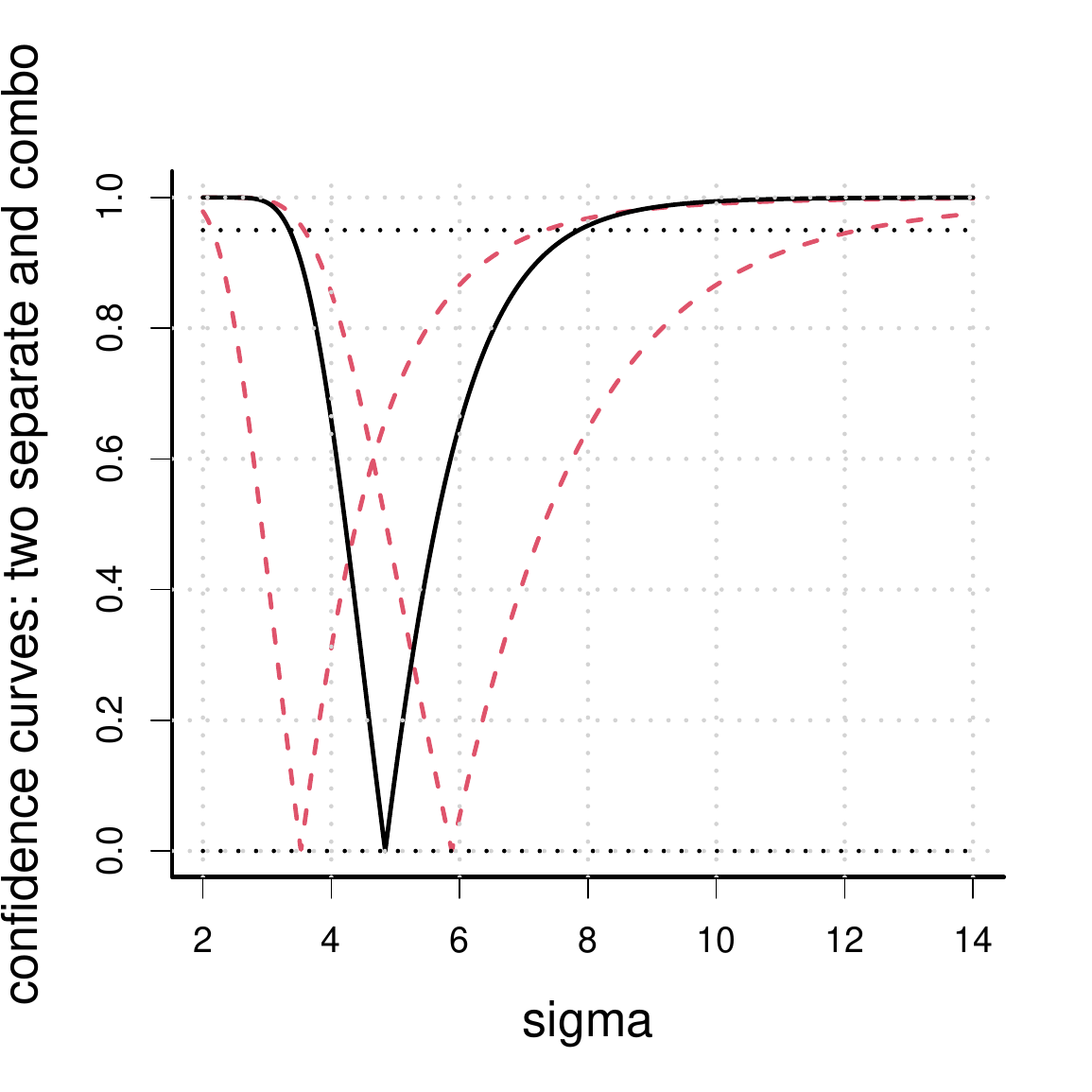} 
\includegraphics[scale=0.35,angle=0]{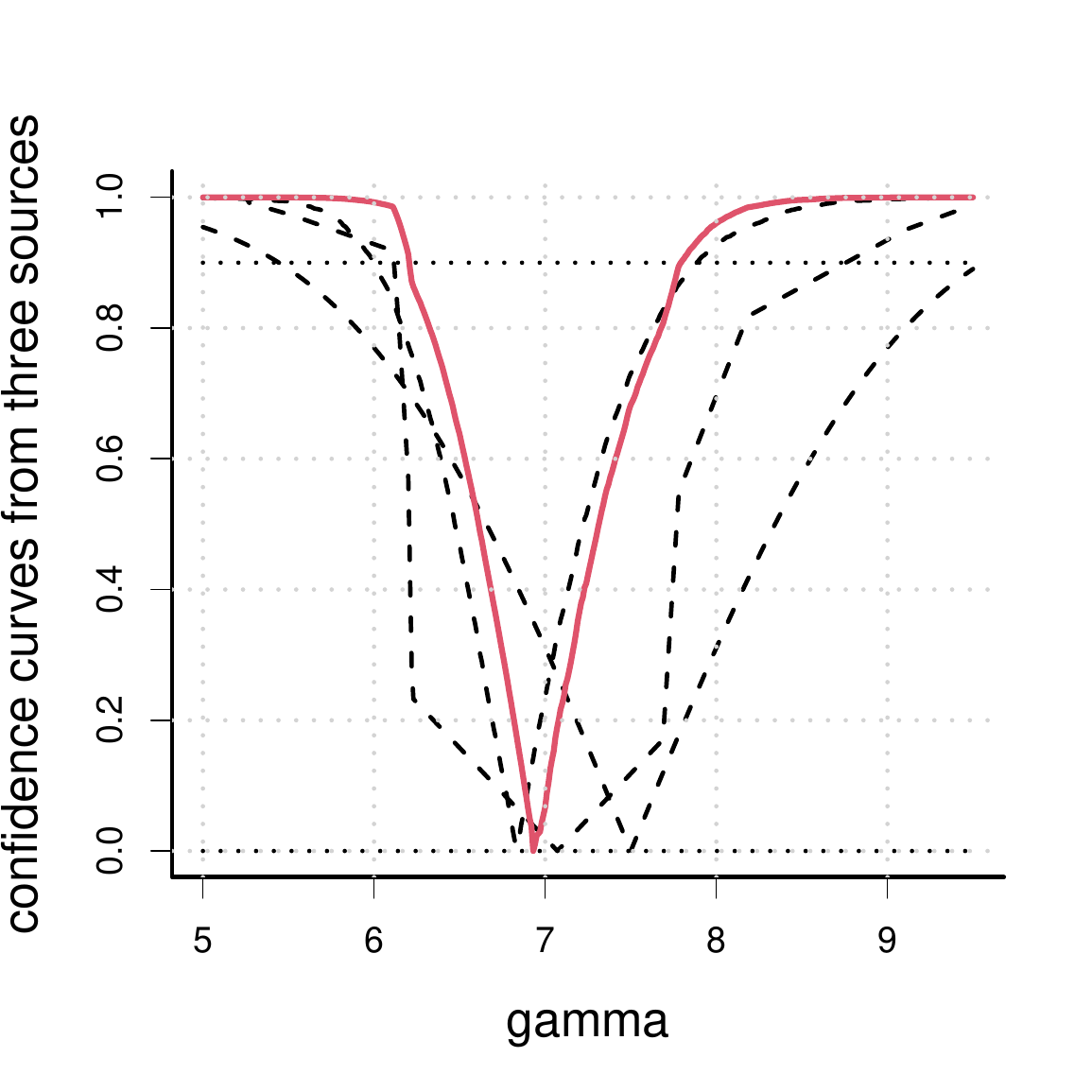} 
\caption{\sl
  Combining information sources:
  Left panel, Illustration A:
  confidence curves $\cc(\sigma)$ for
  tomorrow's standard deviation $\sigma$ (middle curve),
  combining the two estimates 3.33 and 5.55 (outer curves). 
  Right panel, Illustration B:
  confidence curves $\cc(\gamma)$ for the 0.75 quantile
  $\gamma$ of a certain distribution, based on
  three information sources;
  dashed lines for these three, full curve for the fusion.}
\label{figure:combo71}
\end{figure} 

The two illustrations below are partly meant as toy
examples, to see clearly how the information combination
is being handled and fused. It is hoped, though,
that there will be climate science applications
featuring partly the same apparatus. In particular,
the methods enable one to combine frequentist and Bayesian
information, also with e.g.~simulation outcomes via
physical models. One view of the method is that it produces
fused confidence intervals from a collection of individual ones. 
Generalisations of such methods are also able
to combine different predictive distributions;
a case in point here is fusing together
different predictions for `time to frost', in \citet{Cunenetal26}.  

{\it Illustration A:}
Suppose there are two independent estimates of
tomorrow's standard deviation for some variable
of interest, with $(\hatt\sigma_1,\hatt\sigma_2)=(3.33,5.55)$, 
both of the familiar type from normal observations,
with $m=6$ degrees of freedom. We derive from these CDs
$C_j(\sigma)=1-G_m(\hatt\sigma_j^2/\sigma^2)$, for $j=1,2$,
in terms of the $\chi^2_m$ c.d.f. The II-CC-FF method
then gives the confidence curves $\cc^*(\sigma)$
plotted in Figure \ref{figure:combo71} (left panel).
In this fashion the two separate 95 percent confidence intervals,
of $[2.15,7.33]$ and $[3.58,12.22]$, is fused into
the both-sources one of $[3.35,7.85]$.
In this particular case exact inference may also be
carried out, yielding an optimal confidence curve
quite close to the $\cc^*(\sigma)$ one, the latter
having been constructed from the generic II-CC-FF machinery,
without needing particular distribution results from
functions of normal data. 

{\it Illustration B:}
Assume there are three sources carrying
information about $\gamma=F^{-1}(0.75)$, the upper quartile
of some distribution studied. Source 1 has $n_1=20$
data points, believed to be normal, yielding confidence
for the appropriate $\gamma=\xi+0.675\,\sigma$.
Source 2 has $n_2=20$ other data points,
with no parametric assumptions,
giving a nonparametric CD for the quantile, being more wiggly. 
Source 3 is a Bayesian one, expressed in terms
of a normal $(7.50,1.25^2)$. 
Figure \ref{figure:combo71} (right panel)
displays these three separate confidence curves
$\cc_j(\gamma)$, with the nonparametric one being
wiggly. The full middle curve is the result of the
II-CC-FF fusion. The individual 90 percent
intervals are $[6.01,7.88]$, $[6.12,8.75]$, $[5.45,9.50]$,
and when fused together yielding the combined-information
tighter 90 percent interval $[6.21, 7.79]$.

\section{Bolt from heaven: Assessing probabilities for extreme scenarios}   
\label{section:Bolt} 

On 31 May 2008, Usain Bolt burst upon us, with his first 
world record, 9.72 seconds for the 100 meter sprint
(the previous was Powell's 19.74, September 2007).
How surprised were we? 

\begin{figure}[h]
\centering
\includegraphics[scale=0.35,angle=0]{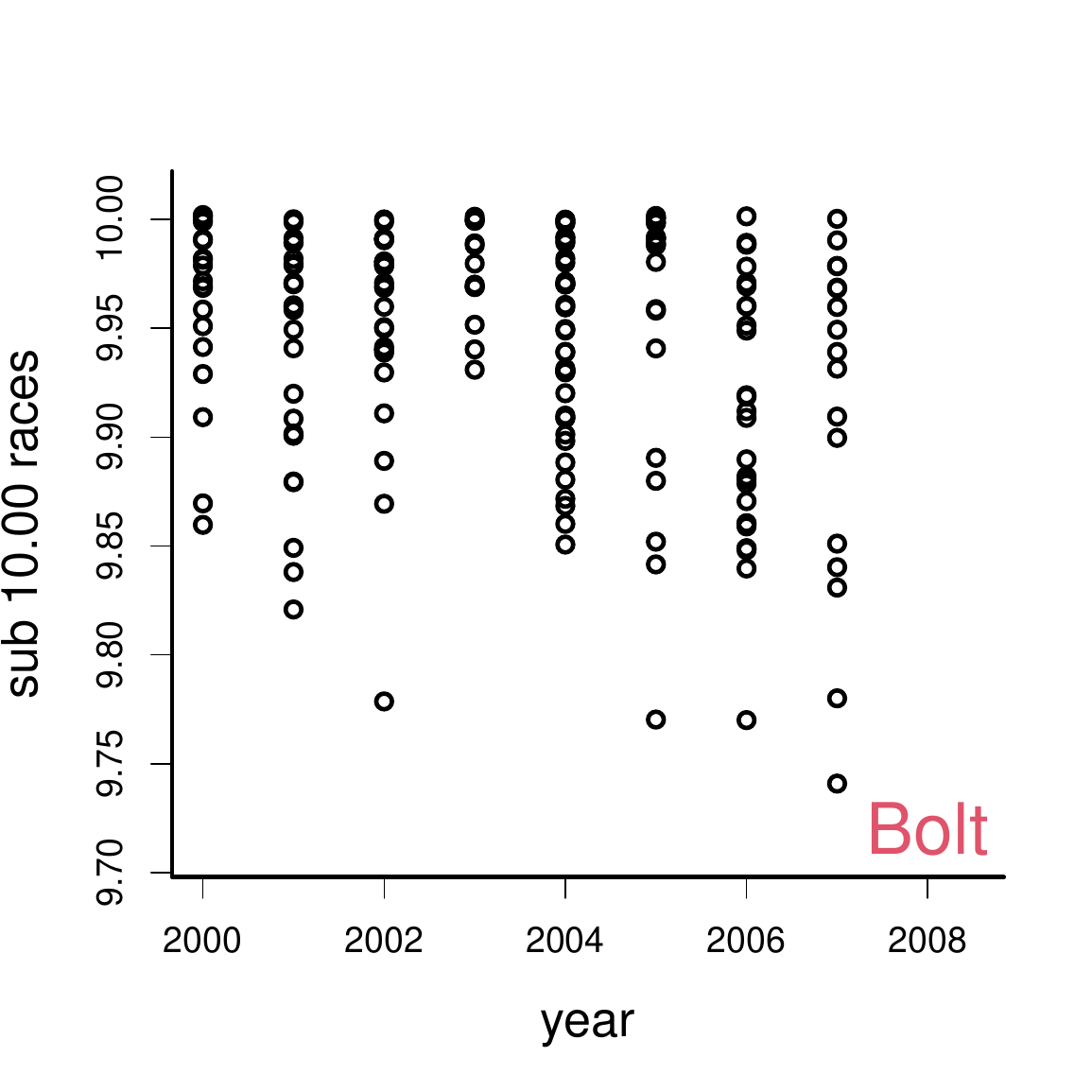}
\includegraphics[scale=0.35,angle=0]{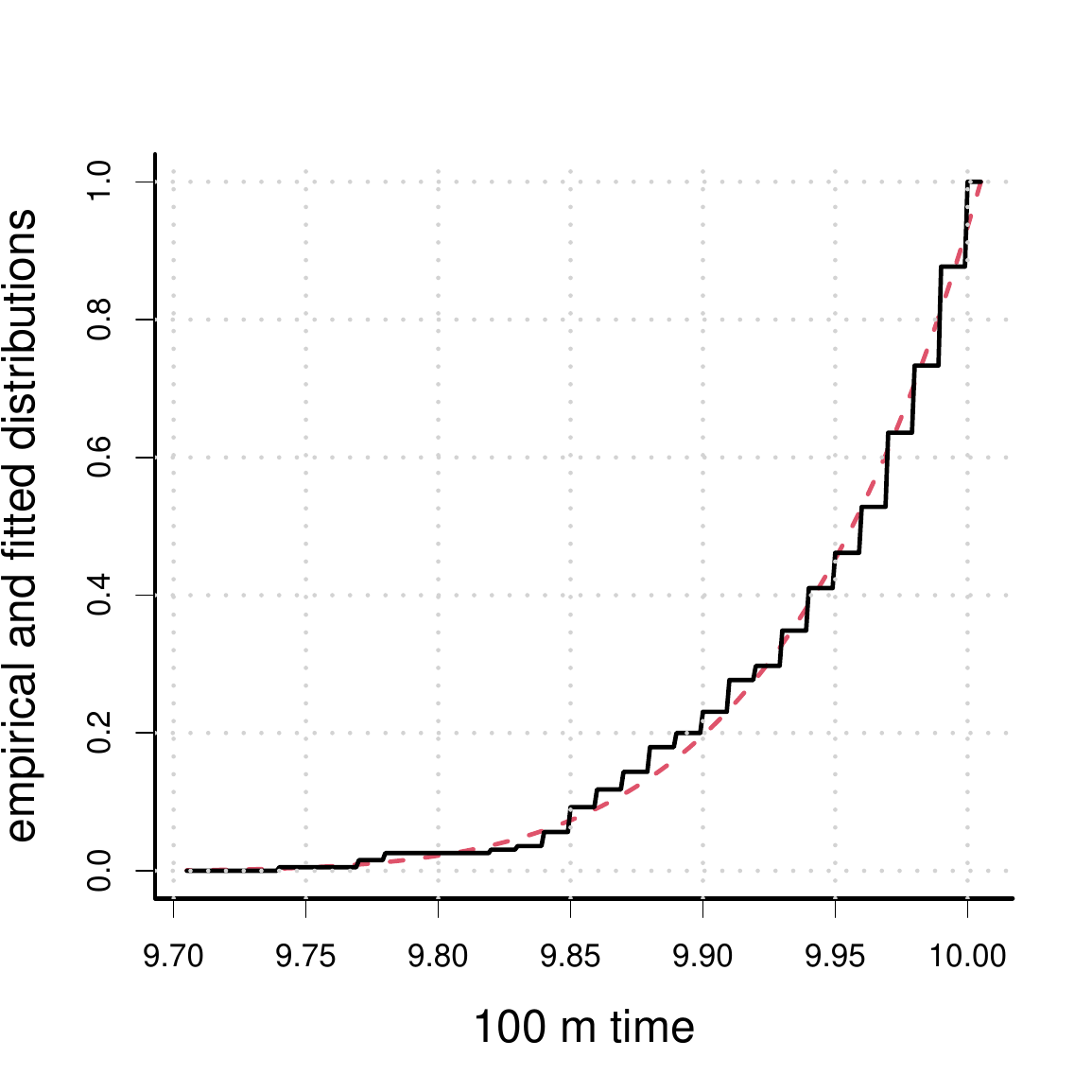}
\caption{\sl
Left panel: all the 195 sub-10.00 races achieved 
during the eight seasons 2000 to 2007, along with the new word 
record 9.72 ran by Bolt in May 2018. 
Right panel: the empirical distribution function (black, rugged)
for these 195 races, along with the fitted two-parameter
distribution from extreme values theory.}
\label{figure:bolt11}
\end{figure}

In various domains one experiences `shocks' like this,
as with the climate sciences: something almost-not-heard-of
happens. There is a need to understand and to assess
just how shocking such events are, including the uncertainty
involved when estimating pertinent probabilities;
methods used and lessons learned form such studies are
also relevant for prediction purposes.

We allow ourselves, therefore, to look into the Bolt event,
trusting that related methods, perhaps with modifications
or extensions, might be used also when assessing
extreme weather, extreme changes in climatic trends, etc. 
To approach the World Record question, along with those which followed 
as Bolt did 9.69 (August 2008) and then
9.58 (August 2009), we compare the 9.72 performance with
the $n=195$ sub-10.00 races of 2000--2007;
see Figure \ref{figure:bolt11} (left panel). 
To readily access a body of literature on extreme values theory,
see e.g.~\citet*{Embrechtsetal97}, we transform these 
race times $r_i$ to $y_i=10.005-r_i$. Such theory predicts 
that the $y_i$ should follow the distribution 
\beqn
G(y,a,\sigma)=1-(1-ay/\sigma)^{1/a} \quad {\rm for\ }y>0, 
\eeqn 
for parameters $(a,\sigma)$. This assumes that the phenomenon
studied has remained reasonably stationary during the period
of observation, here that the top level of sprinting has not
undergone serious changes in the period 2000--2007. 
The log-likelihood function takes the form 
\beqn
\ell(a,\sigma)=\sumin \{-\log\sigma+(1/a-1)\log(1-ay_i/\sigma)\}. 
\eeqn
Fitting the model gives maximum likelihood estimates 
$(\hatt a,\hatt\sigma)=(0.1821,0.0701)$ and leads to 
Figure \ref{figure:bolt11} (right panel). 
As we see, the model works very well. 

For a season with $N$ top races, below the Hary threshold 10.00, 
consider now $p=p(a,\sigma,N)=\Pr(\max(Y_1',\ldots,Y_N')\ge y_0)$.
With $N$ being a $\pois(\lambda)$, this probability 
of seeing a race $r$ with $y=10.005-r\ge y_0$, in the course
of a new season, can be shown to be 
\beqn
p=p(a,\sigma)=1-\exp\{-\lambda(1-ay_0/\sigma)^{1/a}\}. 
\eeqn 
Here we use $\lambda=195/8=24.375$, the rate of top races per year. 
For each threshold $y_0$ we may estimate $p(a,\sigma)$. 
With $y_0=10.005-9.72=0.285$, for 31 May 2008, we find
$\hatt p=0.035$; the estimated probability of seeing 
a 9.72 or better in the course of 2008, as judged from 1 January 2008,
was 3.5 percent. 

\begin{figure}[h]
\centering
\includegraphics[scale=0.35,angle=0]{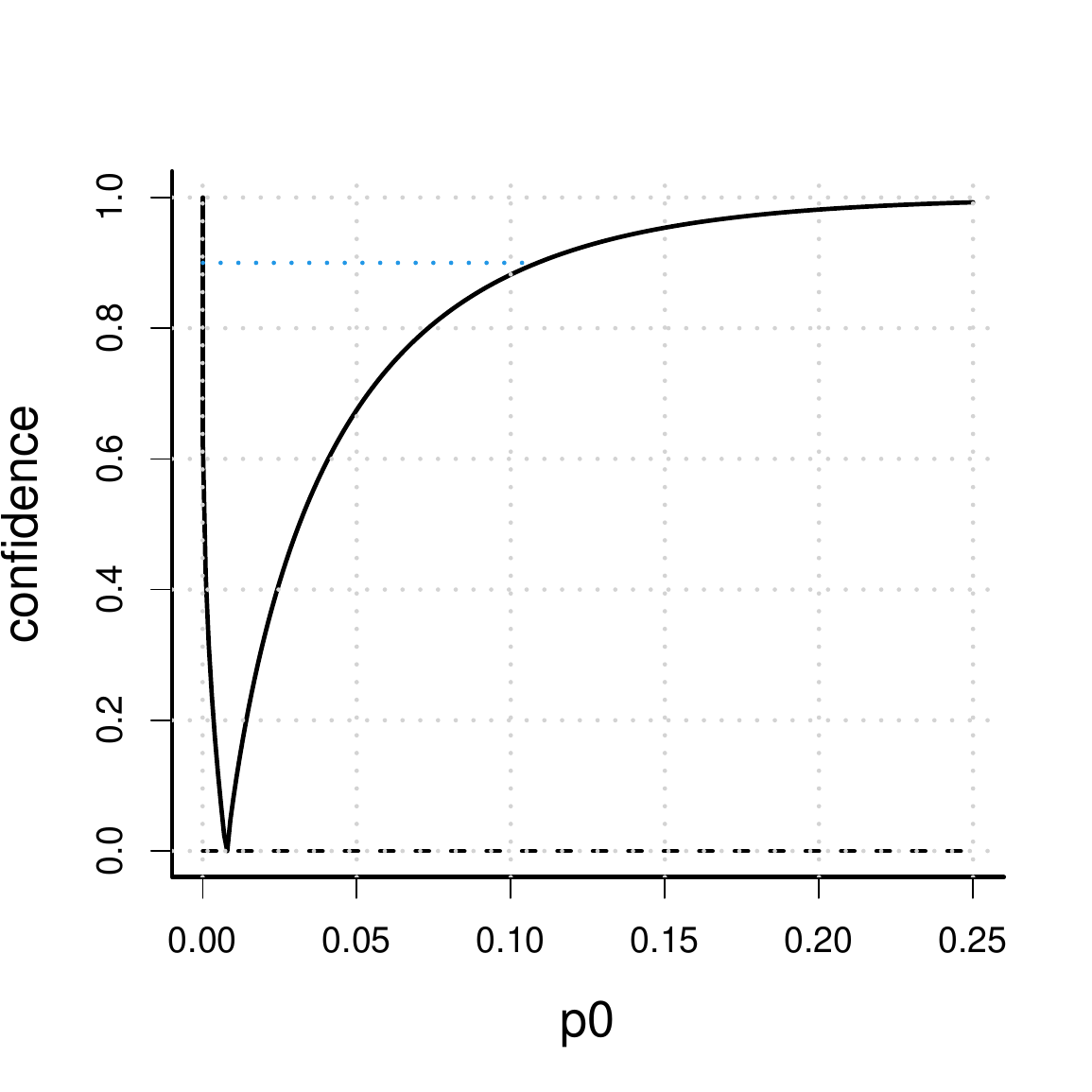}
\includegraphics[scale=0.35,angle=0]{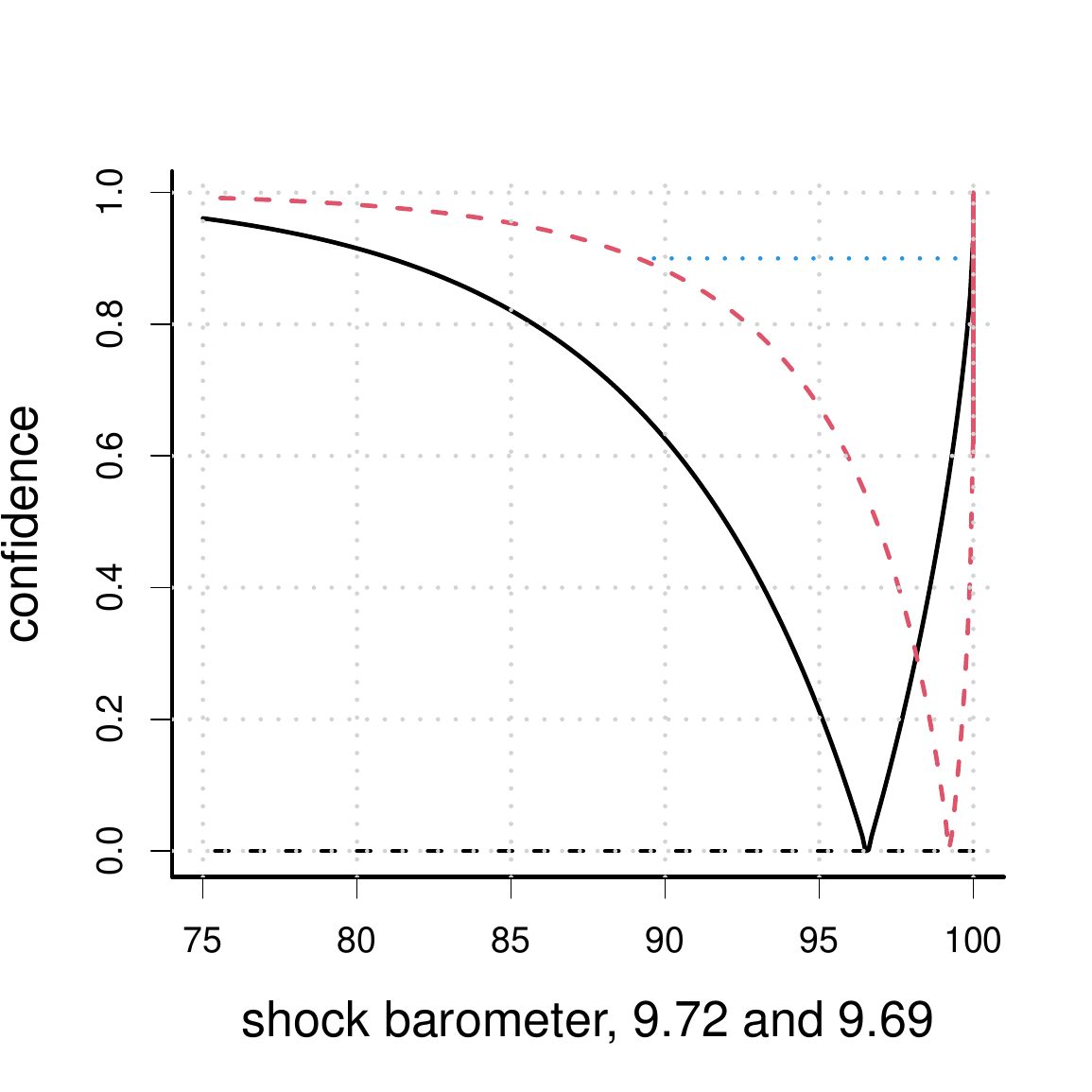}
\caption{\sl 
Left panel: confidence curve for the probability $p$
of seeing a 100 m race with 9.72 or better, as judged
by the start of the 2008 season. The point estimate is 
3.5 percent, but the distribution is rather skewed; 
the 90 percent interval for $p$ is $[0,18.9]$ 
on the percentage scale.
Right panel: the $p$ probability transformed to 
the shock barometer scale $100(1-p)$, with confidence curves
both for 9.72 and 9.69; we were even more shocked 
by his Olympics 2008 race, with $p$ estimated at 0.7 percent,
90 percent confidence interval $[0,10.8]$ percent, 
and shock barometer 99.2. His Berlin 2009 race of 9.58 
is measured to be perfectly Beamonesque, on this scale.}
\label{figure:bolt21}
\end{figure}

The delta method for assessing the variability of 
$\hatt p=p(\hatt a,\hatt\sigma)$ does not work so well here, 
even though $(\hatt a,\hatt\sigma)$ is approximately binormally
distributed. In spite of the sample size $n=195$, the 
function $p(a,\sigma)$ is not well approximated by a linear 
function around the maximum likelihood position. What works better is 
the Wilks theorem and the associated CD methods,
see \citet[Ch.~3]{SchwederHjort16}. 
This requires computation of the log-likelihood profile function 
\beqn
\ell_\prof(p_0)=\max\{\ell(a,\sigma)\colon p(a,\sigma)=p_0\}
\eeqn  
for a grid of $p_0$ values. To this end, one shows first that
$p(a,\sigma)=p_0$ entails $\sigma=aw /( 1 - (\alpha_0/\lambda)^a)$,
with $\alpha_0=-\log(1-p_0)$, leading to an easier one-dimensional 
optimisation problem, for each $p_0$. 
Computing the log-likelihood profile and using the Wilks theorem
recipe, one finds the confidence curve $\cc(p_0)$, 
as with Figure \ref{figure:bolt21} (left panel). 
On the percentage scale, the point estimate is 3.4 and 
the 90 percent interval is $[0,18.9]$, with a veritable skewness.
Similar analyses may be carried out for for Bolt's 2008 Olympics race 
of 9.69. Transforming  estimates and confidence to the 
shock barometer scale of $100(1-p)$, we get 
Figure \ref{figure:bolt21} (right panel). 
His 9.58 in Berlin August 2009 really shattered the scale, being 
very close to being unbelievable, as seen from January 2008
(but after that we had shifted our scales of expectation). 

%

\section{Concluding remarks} 
\label{section:concluding} 

The list below can certainly be extended, with
different prior and posterior views of the inherent
fruitfulness of the various methods envisaged.
A `let us explore' attitude here would be welcomed,
both for the theory and for the usefulness in given
applied projects. 

\smallskip
{\it A. Bayes.}
The use of Bayesian statistics, from the use of careful
prior elicitation and clarification for components of
complex models to uncertainty quantification,
is prevalent in climate statistics.
The prior probability that yet more fruitful
methodological and explorative work in this domain can be
accomplished exceeds 95 percent.
The \citet*{LerchThordis17} article is a case in point,
with application to uncertainty quantification for
ensembles of forecasts.
The \citet{Hasselmann98} article was in fact among the
first Bayesian contributions when addressing detection
and attributions problems.

Time series of course abound in climate studies,
where there is a good catalogue of
well-understood models with well-working tools,
see e.g.~\citet[Ch.~10]{Storch02}. Here there would be
room also for Bayesian nonparametrics, where the
autocorrelation structure can be modelled via
nonparametric envelopes around the most well-used
parametric models. Machinery for such, involving
also setting Bayesian priors for the spectral
distribution of stationary series, is outlined in
\citet[Section 4]{Hjort10chapter}.
The skiing days time series data analysed
in Section \ref{section:bjoernholt}, via a four-parameter
model, can very well be worked with using tools
from Bayesian nonparametrics, starting with a more
flexible model for the trend function itself
than the forced-to-be-linear $a+bx$. 

\smallskip
{\it B. The Focused Information Criterion.}
As briefly pointed to in connection with the AIC
of (\ref{eq:aic}), the FIC machinery of
\citet{ClaeskensHjort03, ClaeskensHjort08} should be
a fruitful venue for comparing and ranking candidate
models for parts of climate statistics. The core idea
is that different foci, i.e.~different goals with
the full analysis, should lead to different rankings
and indeed perhaps different `winners'. A good model
for predicting ten years ahead is not necessarily
a good model for assessing changes in correlation structure,
and vice versa.

For various applications there would be several such
`very good models', inviting also focused averages
of estimates or predictions to be constructed;
see the references pointed to. There is a sometimes
subtle difference between (i) we construct as good
estimates and predictions as we can and (ii) we
work with a good model to help us identify,
assess, and interpret their key parameters. Model averaging,
whether Bayesian or frequentist, could help with (i),
but not necessarily with~(ii). 

\smallskip
{\it C. Long-range and intermediate-range time series.}
There is a certain debate regarding the sometimes long-memory
properties of long historical temperature series,
as discussed in e.g.~\citet*{Dagsvik20}. It is not easy
to give a very clear `yes' or `no' to the question
`is there long-range dependence in my long time series'.
Model comparison tools of the FIC kind should be able
to sort out such questions better. It ought also to
be a fruitful endeavour to used Bayesian nonparametrics
tools for the spectral distributions involved,
to allow a flexible interpretable framework encompassing
both short-, intermediate-, and long-range models. 

\smallskip
{\it D. Change points and regime shifts.}
There is a rich and growing literature on change points,
for both theory and practice, where applications abound.
Notably, a change point does not have to be the
basic `response changes from one level to another',
but could relate to certain components inside bigger
models for bigger questions. An example could be
a climate process where both the overall level and
the overall variance stay about the same, but where
the internal autocorrelation structure changes.
Questions worked with in Sections \ref{section:future} 
and \ref{section:HjortKola} are also of this type;
there could be a change in the derivative,
or in one of the regression coefficients, for systems
unfolding over longer time periods. Change point identification
methods are used in a plethora of application domains,
including as different topics as literature
(when did Author B take over from Author A,
see \citet*{CunenHermansenHjort18})
and peace and conflict research
(has the world become more peaceful, \citet*{CunenHjortNygaard20}). 

\smallskip
{\it E. Interplay and causality.}
As we saw in Section \ref{section:HjortKola},
issues of climate have sometimes rich and complex
relations with other evolving phenomena, from
people's lives to economy and biology. There is room
for more theory, and more exploration, regarding
both the joint modelling of climate Y and consequences X
or sometimes the other way around, where X influences Y
in a causal manner. This points to yet another
complex theme, that of statistical causality,
which famously is more difficult to pinpoint and assess
than mere correlations and intervariability.
From a growing literature in such directions,
methods developed in \citet{Silvaetal21}, applied to
detecting climate teleconnections, appears fruitful 


\section*{Acknowledgements}

The author is grateful for the opportunity to
meet and fruitfully discuss modern statistical methods
with esteemed climate scientists at the Hamburg Hasselmann
Legacy conference in November 2024. Comments from
Hans von Storch led to an improved presentation
of this chapter. The positive welcoming atmosphere
during that conference also enabled him to give
a short extra impromptu talk on
``Wann, warum und wie Hamburg zu einer
friedlichen Stadt wurde'',
in connection with Professor von Storch's PhD cavalcade day. 


\begin{small} 

\bibliographystyle{biometrika}
\bibliography{diverse_bibliography2025}

\end{small} 

\end{document}